\documentclass[10pt, journal, oneside, twocolumn]{IEEEtran}

\usepackage{hhline}
\usepackage{pdfpages}
\usepackage{adjustbox}  %
\usepackage{amsmath}    %
\usepackage{float}      %
\usepackage{graphicx}   %
\usepackage{hyperref}   %
\usepackage{fancyhdr}   %
\usepackage{footnote}   %
\usepackage{cleveref}
\usepackage{amsfonts}
\usepackage{multirow}
\usepackage{algorithm}
\usepackage[noend]{algpseudocode} %
\usepackage{pgf-umlsd}
\usepackage{circuitikz}
\usepackage{tikz}
 \usepackage{amssymb} 
 \usetikzlibrary{shapes.geometric, arrows, positioning,shapes.multipart,fit,backgrounds,calc,matrix}
 \usepackage{booktabs}
 \usepackage{tabularx}
 \usepackage{threeparttable}
 \usepackage{svg}
\usepackage{amsmath}
\usepackage{tikz-uml}        %
\usepackage{fontawesome5} %
\usepackage{units}
\usepackage[T1]{fontenc}
\usepackage[utf8]{inputenc}

\renewcommand{\newline}{\vspace{0.3cm}}

\newcommand{\mr}[1]{\mathrm{#1}}
\newcommand{\argmax}{\operatorname*{arg\,max}}

\newcommand{\Score}{\operatorname{Score}}
\newcommand{\AssignedTasks}{\operatorname{AssignedTasks}}
\newcommand{\LockedVehicles}{\operatorname{LockedVehicles}}
\newcommand{\EligibleDepartures}{\operatorname{EligibleDepartures}}
\newcommand{\GreedyDispatches}{\operatorname{GreedyDispatches}}

\newcommand{\MultiDispatch}{\operatorname{MultiDispatch}}

\usepackage[backend=biber,style=ieee]{biblatex}
\def\BibTeX{{\rm B\kern-.05em{\sc i\kern-.025em b}\kern-.08em
    T\kern-.1667em\lower.7ex\hbox{E}\kern-.125emX}}

\DeclareSourcemap{
  \maps[datatype=bibtex]{
    \map{
      \step[fieldsource=entrykey, match=VehlhaberSalazar2023b, final]
      \step[fieldset=abstract, null]
    }
  }
}

\apptocmd{\sloppy}{\hbadness 10000\relax}{}{}

\tikzset{
  every node/.style = {font=\large, align=center},
  arrow/.style     = {-{Stealth[length=3mm,width=2mm]}, line width=0.9pt, font = \small},
  shortline/.style = {line width=0.9pt},
  state/.style     = {draw, rounded corners=8pt, line width=0.9pt,
                      fill=white, minimum width=2.8cm, minimum height=1cm},
  move/.style      = {draw, rounded corners=3pt, line width=0.9pt,
                      fill=gray!30, minimum width=2.6cm, minimum height=1cm},
  task/.style      = {draw, rounded corners=3pt, line width=0.9pt,
                      fill=gray!50, minimum width=2.6cm, minimum height=1cm},
  band/.style      = {draw= black, dashed, rounded corners=2pt, fill=gray!10},
  bandlabel/.style = {font=\Large, inner sep=1pt, align=left},
    decision/.style = {
    draw,
    regular polygon,
    regular polygon sides=4,
    shape border rotate=45,
    minimum width=0.7cm,
    minimum height=0.7cm,
    align=center,
    fill=white,
    line width=0.9pt
  },
}  
\newcommand{\vsp}{-5pt} %

\begin{document}

\title{Optimizing Airport Electric Ground Support Fleets with Digital-Twin-Guided Predictive Control}

\author{Juan Pablo Bertucci$^{1}$,
Mauro Salazar$^{1}$,
and Theo Hofman$^{1}$%
\thanks{This publication is part of the project GTD-Elektrifikatie, made possible in part by the Ministry of Economic Affairs and Climate Policy of the Netherlands.}%

\thanks{$^{1}$Department of Mechanical Engineering, Control Systems Technology,
        Eindhoven University of Technology, 5600 MB  Eindhoven, The Netherlands
        {\tt\small \{j.p.bertucci\}@tue.nl}}%
}

\maketitle

\begin{abstract}
Electrifying airport ground operations is a key step toward reducing aviation-related emissions, but it makes task assignment and charging coordination strongly coupled in real time. 
This paper develops a digital twin of airport ground handling operations for electric ground-support vehicles (EGSVs), combining an agent-based model of airport operations with live flight-schedule updates and vehicle-energy dynamics. 
On this platform, we compare three control strategies for EGSV dispatching and charging: a rule-based heuristic, a rolling-horizon model predictive controller (MPC), and a Monte Carlo tree search (MCTS) controller.
The controllers are evaluated on three representative Schiphol Airport operating days and for multiple fleet sizes. The results show a clear trade-off between delays and energy use. 
Across all tested scenarios, MCTS yields the lowest average flight delay, with a pooled delay of 4.09 min, improving on the heuristic and MPC controllers by 64.7\% and 31.5\%, but at an increased energy consumption of 1\% and 0.3\%, respectively.
The MPC and MCTS controllers coordinate charging more effectively throughout the day and reduce the peak charging infrastructure required by 6.0 and 4.7 charger units, respectively.
Additionally, we find more vehicles can result in lower overall energy consumption and lower charging infrastructure requirements for all controllers.
These findings highlight the value of digital twins not only as offline design evaluation tools, but also as practical environments for real-time control policies in electrified airport operations.
\end{abstract}

\section{Introduction}\label{sec:intro}
\begin{figure}[t]
\centering
\resizebox{\linewidth}{!}{%
\tikzset{
  >=Stealth,
  font=\small,
  block/.style={draw,rounded corners,minimum width=3.6cm,minimum height=1.1cm,align=center,fill=white},
  splitblock/.style={block,rectangle split,rectangle split parts=#1,align=center},
  icon/.style={font=\large,inner sep=1pt},
  flow/schedule/.style={-Latex,very thick,blue},
  flow/assign/.style={-Latex,very thick,black},
  flow/measure/.style={-Latex,thick,gray,densely dashed},
  legend box/.style={draw,rounded corners,fill=white,inner sep=4pt}
}

\begin{tikzpicture}[node distance=1.6cm,scale=0.98,transform shape]

\node (sched) {\faCalendar*};

\node[block,right=1.25cm of sched] (ctrl) {\faCogs~Controller\\(Heur / MPC / MCTS)};

\node[splitblock=4,right=2.4cm of ctrl] (twin)
  {\faProjectDiagram~Airport Digital Twin
   \nodepart{two} \faPlane~Agent: Aircraft
   \nodepart{three} \faShuttleVan~Agent: Apron Vehicle
   \nodepart{four} \faChargingStation~Agent: Charger Station};

\node[right=2.0cm of twin] (out) {\faWaveSquare};

\draw[flow/schedule] (sched) -- node[above]{${I}(t)$} (ctrl);
\draw[flow/assign]   (ctrl)  -- node[above]{${X}(t)$} (twin);
\draw[flow/measure]  (twin)  -- node[above]{KPI(t)} (out);
\path (twin.south) |- ++(0,-0.5) coordinate(mid) -| (ctrl.south);
\draw[flow/measure] (twin.south) |- ++(0,-0.5) -| (ctrl.south);
\node[below] at ($(mid)!0.5!(ctrl.south)$) {$s(t)$};

\node[legend box,below=2.0cm of ctrl,align=left] (legend) {
  \begin{tabular}{@{}ll@{}}
    {\tikz\draw[flow/schedule] (0,0)--++(8mm,0);} & API input \\
    {\tikz\draw[flow/assign]   (0,0)--++(8mm,0);} & controller outputs \\
    {\tikz\draw[flow/measure]  (0,0)--++(8mm,0);} & DT outputs \\
  \end{tabular}
};

\end{tikzpicture}

  
}%
\caption{Sketch of the problem setup. Task schedules ${I}(t)$ are drawn from the Schiphol API and fed to the controller, which provides the assignment of tasks ${X}(t)$ to the airport DT at every time step. The twin outputs state $s(t)$ feedback and performance metrics $KPI(t)$.}
\label{fig:intro}
\end{figure}
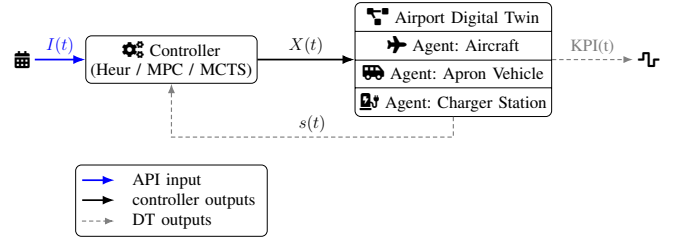

Aviation emissions pose a significant environmental challenge, with ground support operation (taxiing, push-back, passenger and baggage transport, etc.) emissions constituting up to 10\% of an airport's total carbon footprint~\cite{catapultReport}. At Copenhagen Airport, these emissions represent between 5-9\%~\cite{WINTHER2015} while at Bologna Airport they account for 9.3\% of total emissions~\cite{POSTORINO2014}.
To address this issue, airports are transitioning to carbon-neutral vehicle fleets. For example, Schiphol Airport aims to achieve an emission-free ground fleet by 2030~\cite{schipholemmissie}. 
These fleets comprise heterogeneous Ground Support Vehicles (GSVs), including passenger buses, baggage tractors, and fuel trucks, whose coordinated operation is essential for efficient and punctual Ground Handling Operations (GHO)
As these fleets become electric, the joint scheduling of tasks and charging activities for Electric GSVs (EGSVs) becomes critical to avoid operational delays.
In this paper, we leverage advanced simulation techniques to provide a Digital Twin (DT) platform, as shown in Fig.~\ref{fig:intro}, for the control of EGSVs and employ heuristic and predictive control approaches to analyze control impacts on design, efficiency and performance metrics.
This work advances the Intelligent Transportation Systems (ITS) literature by demonstrating the value of DTs as platforms for developing, testing and deploying advanced control strategies in complex, time-sensitive environments.

\textit{Literature review:}
Several studies have focused on the optimization of airport GHOs using mathematical optimization, heuristics and simulation-based approaches.
Airport-specific models have addressed apron vehicle assignment, dispatching and scheduling under operational constraints using mathematical optimization, showing travel distance reductions, improved resource usage and increased schedule robustness~\cite{Andreatta2014,Tabares2021,GuimaransPadron2022}.
GHOs have also been modeled as a large-scale vehicle-routing and scheduling problem across multiple service types, where realistic instances are tackled using gradient-free optimization (large neighborhood search and genetic algorithms)~\cite{Zhou2023,Yao2024}.
These studies show the value of optimization for airport operations, but they typically focus on vehicle scheduling and resource allocation without explicitly considering the additional constraints introduced by fleet electrification.
A second stream of literature focuses explicitly EGSVs.
In the broader EV-routing literature, several models extend routing problems to include charging stops, time-windows and electric vehicle-specific constraints~\cite{Brevoord2021,Zoutendijk2023}.
Charger availability and partial recharging decisions have also been modeled explicitly, showing that such features can substantially affect feasibility and cost~\cite{KeskinCatay2016,Yu2023}.
Furthermore, offline optimization studies on electric vehicles have shown that fleet sizing, charging-task scheduling and infrastructure design are tightly coupled in electric fleet operations~\cite{PaparellaHofmanEtAl2024,PaparellaChauhanEtAl2024,BertucciSalazarEtAl2025}.
Within airport-specific applications, recent studies formulate the scheduling of electric or mixed ground-handling fleets with charging constraints as bi-objective, decomposition-based or cooperative scheduling problems~\cite{Bao2023,Fu2025,Tang2025,ZoutendijkMitici2024}, while stochastic dynamic planning models have been proposed for electric towing vehicles under uncertain aircraft arrival and departure times~\cite{vanOosteromMitici2024}.
Beyond ground handling, the electrification of aviation networks more broadly has motivated optimization-based approaches for flight scheduling and energy management~\cite{VehlhaberSalazar2023b,FernandezZapicoIzadiEtAl2026,VehlhaberSalazar2026}, as well as co-design frameworks that jointly optimize fleet size, battery capacity and charging infrastructure~\cite{BertucciRaghuramanEtAl2025,IzadiFernandezZapicoEtAl2026}.
These studies demonstrate the growing relevance of EGSV optimization in airports, but they mainly focus on offline optimization or a single scheduling formulation rather than on online control within a broader airport operational execution environment.
A third stream of literature addresses real-time decision-making for transportation and fleet systems.
In broader mobility contexts, Model Predictive Control (MPC) has been applied to reschedule operational tasks and handle disruptions in an online fashion~\cite{Tsao2019,Cavone2022,FernandezZapicoHofmanEtAl2025,VehlhaberSalazar2026b}.
Heuristic and search-based methods such as Monte Carlo Tree Search (MCTS) are viable alternatives for high-dimensional, stochastic fleets because they strike a balance between solution quality and computational time~\cite{Swiechowski2022MonteApplications}.
Deep learning methods have also shown promising results for large-scale airport ground-handling scheduling~\cite{wu2023neural,Zhou2023}.
However, few studies apply and compare such controller families specifically for heterogeneous airport EGSV fleets while jointly accounting for charging, infrastructure limitations and stochastic execution.
Digital twins have also been explored in airport management for multiple purposes.
Existing studies have used DT concepts for airport information management and operational synchronization~\cite{Conde2022}, while related airport applications have considered simulation-supported operational control and DT-based scheduling for airport vehicles and airside systems~\cite{Noorman2018,Luo2024,Attar2024}.
These studies confirm the value of DTs for airport operations, but they do not focus on the real-time control of heterogeneous EGSV fleets with explicit charging and energy constraints.
To the best of the authors' knowledge, no prior work integrates an airport ground-operations digital twin, explicit EGSV energy and charging modeling, and a comparative evaluation of multiple lookahead-based controller approaches for real-time EGSV fleet management under realistic operating conditions, including infrastructure  design impacts.

\textit{Statement of Contributions:}  
This paper makes three main contributions. 

First, we develop a digital twin of airport ground handling operations that integrates real-time flight-schedule data via the airport’s API, and explicitly models charging, energy consumption and operational constraints of EGSVs. 
Second, we implement and compare three distinct online controller approaches for managing an EGSV fleet, enabling comparison under stochastic scheduling and charging conditions. 
Third, we use the DT platform as an experimental environment to perform a comparative evaluation of these controller approaches and their operational and infrastructure implications under realistic conditions. 
With this, we advance the state-of-the-art in airport ground handling operations for electrified fleets, providing insights into controller performance and infrastructure requirements.

\textit{Organization:} The remainder of the paper is structured as follows: In Section \ref{sec:abm} we describe the DT, in Section \ref{sec:controllers} we develop the controllers to be implemented. In Section \ref{sec:Results}, we present empirical results and performance evaluations derived from applying our proposed models to real-world scenarios. 
We conclude the paper in Section \ref{sec:Conclusion} with the key take-aways from our findings and suggestions for future work.

\section{Digital Twin for Schiphol Airport}\label{sec:abm}
\begin{figure}[t]
\centering
\includegraphics[width=1\linewidth]{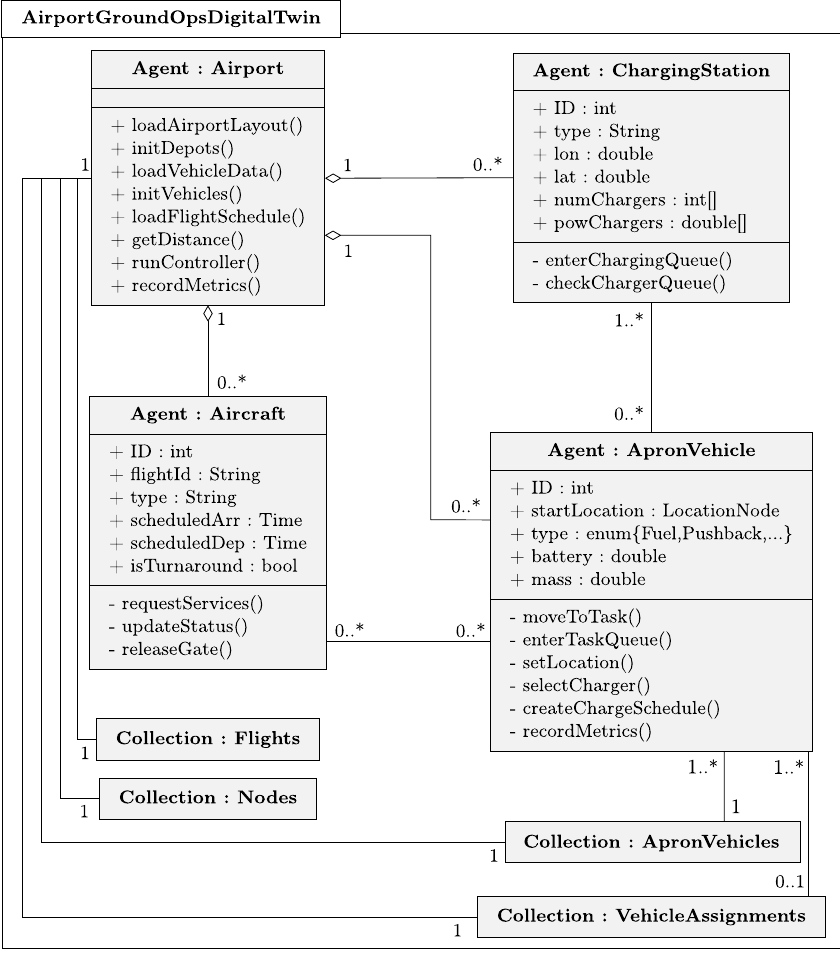}
\caption{UML diagram of the Airport Ground Operations DT, showing the main agents, their attributes, methods, and relationships.}
\label{fig:uml}
\end{figure}

We consider the general problem of scheduling a fleet of electric ground support vehicles (EGSVs) to perform a set of turnaround tasks for a given flight schedule at an airport. The EGSVs are responsible for performing various services such as baggage handling, catering, and passenger transport, which are essential for the timely departure of flights. The scheduling problem involves assigning EGSVs to tasks across the airport's spatial layout while considering constraints such as task precedence, vehicle availability, battery capacities, and charging infrastructure requirements. Additionally, the scheduling must account for uncertainties in flight operations, such as delays and cancellations, which can impact the demand for ground support services. The objective is to minimize flight delays while optimizing energy consumption and charging infrastructure usage. In the following we outline the main agents used to develop a DT of Schiphol Airport. The DT models (i) the airport environment and layout, (ii) EGSV movement and energy use, (iii) aircraft turnaround processes, and (iv) charging infrastructure. It includes four primary agent types: Airport Agents, ApronVehicle Agents, Aircraft Agents, and ChargingStation Agents, each described in the subsequent sections. The main agents and their interactions in the simulation are shown in Fig.~\ref{fig:uml}. An in-depth discussion of each agent and the simulation calculations can be found in~\cite{vincent}.
\begin{figure}[t]%
    \centering
    \includegraphics[width=1\linewidth]{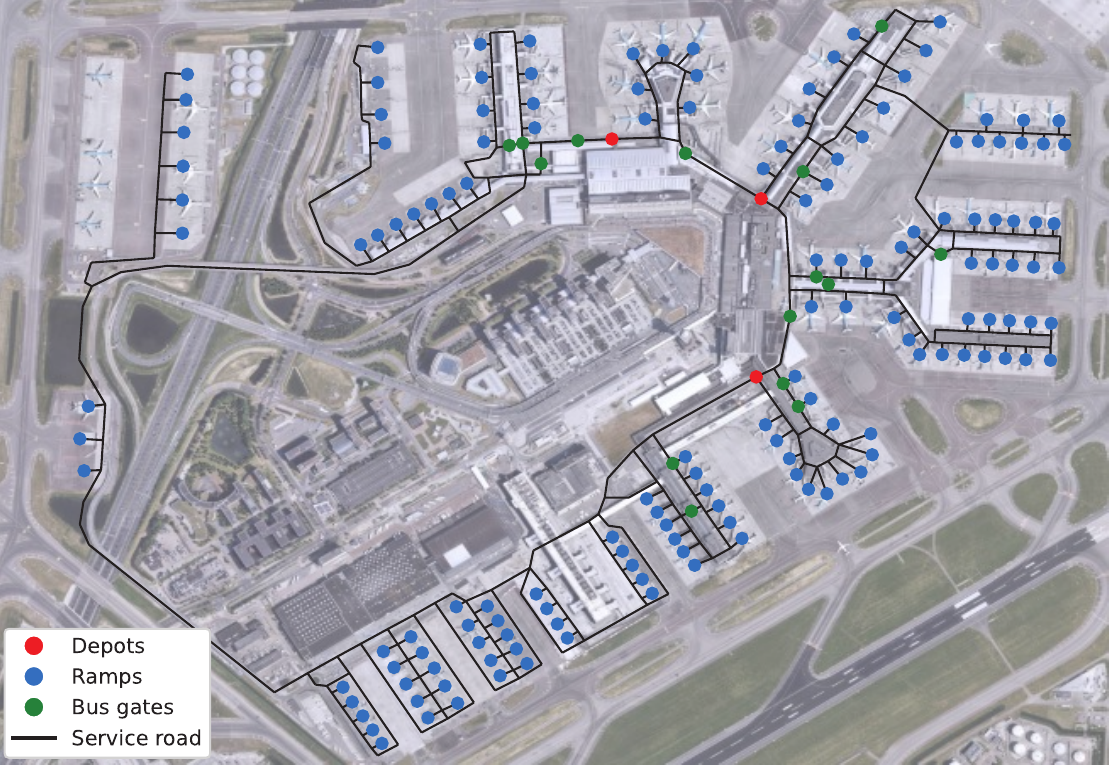}
    \caption{Simplified network of service roads and aprons used for routing in the DT overlayed on aerial view of Schiphol Airport. Depots are indicated in red, aircraft aprons are shown in blue, and passenger transfer stands used by buses are shown in green. Service roads are shown in black. Satellite Image Source: Google Earth.}
    \label{fig:layout}
\end{figure}
\subsection{Airport Agent}
The Airport agent defines the simulation environment in which all other agents are initialised and interact. It stores the network of transitable service roads, and on startup, flight-schedule data from the Schiphol API is used to generate Aircraft agent arrivals.
Fig.~\ref{fig:layout} depicts the simplified layout of the primary service roads at Schiphol Airport, constructed from OpenStreetMap data~\cite{osm} and official documentation~\cite{schipholDocs}. The layout consists of eight piers, two apron areas, and 142 aircraft stands.
The DT relies on flight-schedule data obtained through the Schiphol Flight API, including scheduled and actual arrival/departure times and apron stand locations, to generate a complete daily schedule~\cite{SchipholAPIs}. Aircraft models are obtained by matching their registration code to open databases~\cite{IATA}, and this classification, together with operational manuals, provides typical turnaround-task durations. (A table detailing the turnaround, uplift, and other tasks durations is given in Annex~\ref{annex1}).
The DT incorporates two types of disruptions: temporal and spatial. Temporal disruptions correspond to discrepancies between scheduled and actual arrival or departure times, while spatial disruptions reflect changes in aircraft apron-stand assignment.
Each Aircraft agent generates service tasks that must be completed before pushback, and these tasks form the demand that ApronVehicle agents respond to.

\subsection{ApronVehicle Agent}\label{apronvehicle} 
\begin{figure}[t]
\centering
\resizebox{\linewidth}{!}{%
\includegraphics{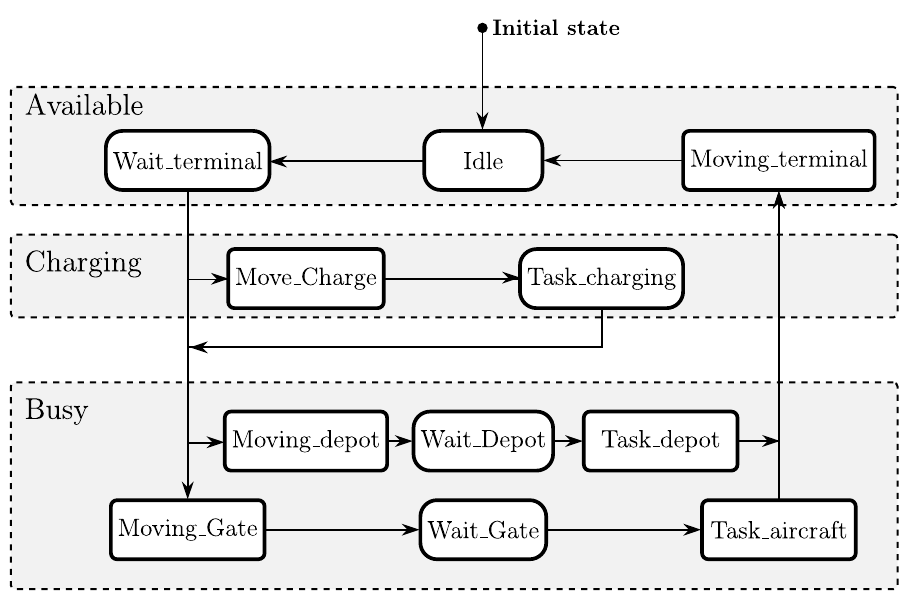} 
} %
\caption{State chart of the ApronVehicle agent. Vehicles are initially available at the terminals, and can move to fulfill a task at a depot or to a gate. Once at the location, it waits until it can perform its task, and after the service time heads back to the depot.}
\label{fig:apronstate}
\end{figure}

ApronVehicle agents represent individual EGSVs responsible for executing specific service tasks at aircraft stands. Each agent has a type parameter that determines its operational role.
At the start of each day, ApronVehicle agents are located at one of three depot sites. They move between stands and depots according to task assignments, traversing the statechart shown in Fig.~\ref{fig:apronstate}, and return to a depot at the end of operations. Each ApronVehicle maintains a queue of assigned tasks, represented as an ordered list of aircraft identifiers.
The ApronVehicle agent uses a physics-based model for driving energy consumption. The instantaneous power is
\begin{equation}
    P^{\mathrm{drv}} = \frac{F_x\,\nu}{\eta},
\end{equation}
where \(\nu\) is vehicle speed, \(\eta\) is the powertrain efficiency, and the longitudinal force \(F_x\) is
\begin{equation}\label{eq:powerFx}
F_x = \mu\,m\,g + \tfrac12\,\rho\,C_d\,A\,(\nu)^2 + m\,a_x.
\end{equation}
Here \(\mu\) is the rolling resistance coefficient, \(m\) vehicle mass, \(g\) gravitational acceleration, \(\rho\) air density, \(C_d\) drag coefficient, \(A\) frontal area, and \(a_x\) acceleration. Acceleration \(a_x\) is computed in-simulation considering the current \(\nu\) and target \(\nu_\mr{tgt}\) speed of the agent. 
This target speed is given by the allowable speed on service roads. 
For vehicles with variable load (baggage tractors, buses), \(m = m_v + m_c\) where \(m_v\) is the vehicle mass \(m_c\) is the cargo/passenger mass.
Energy use during non-driving tasks is assumed negligible except for hydrant dispensers and pushback tugs, which will use instantaneous power $P^{\mathrm{tsk}}$ for the duration of the task.
A list of the technical specifications and assumptions for the ESGVs is provided in Annex~\ref{annex2}.
\subsection{Aircraft Agent}
Aircraft agents represent the flights that require service from ApronVehicle agents. Each Aircraft agent is assigned a stand, type, scheduled arrival/departure time, and passenger bus requirement, according to the flight-schedule data. The stand assignment and schedule are updated according to live API information. Aircraft type and registration data determine task durations and the corresponding ApronVehicle load requirements following the specifications in Annex~\ref{annex1}. The aircraft-specific parameter values reported in Table IV were obtained from manufacturer airport planning manuals.

\begin{figure}[t]
\centering
\resizebox{0.8\linewidth}{!}{%
\includegraphics{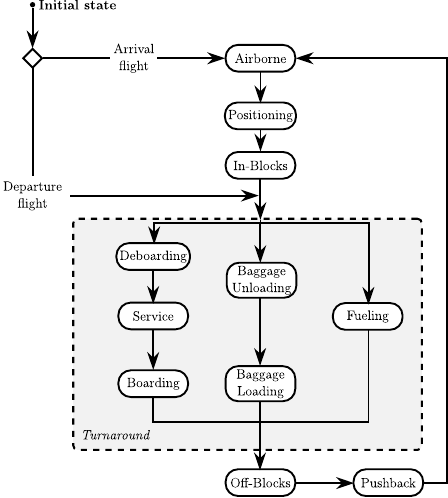} 
} %
\caption{State chart of the Aircraft agent. An aircraft goes through various states during its turnaround process, starting from positioning to pushback. Service tasks are performed in parallel during the servicing state.}
\label{fig:airstate}
\end{figure}

The state transitions of the Aircraft agent, shown in Fig.~\ref{fig:airstate}, represent aircraft turnaround phases, including positioning, deboarding, servicing, boarding, and pushback. These states update dynamically in response to temporal disruptions (delays) and spatial disruptions (stand changes) within the DT environment.

\subsection{ChargingStation Agent}
Charging Station agents represent individual chargers with fixed locations and power capacities. A charger begins in the \textit{Available} state. 
When an ApronVehicle receives a charging task, it requests a charger. If available, the charger transitions to \textit{Reserved} and the vehicle moves toward its location. Upon arrival, the state changes to \textit{Charging} and remains so until the vehicle is fully charged or until the controller stops charging. The charger then returns to \textit{Available}. If no charger is available, the ApronVehicle queues or reroutes according to the controller policy.

\subsection{Vehicle Control}\label{sec:abm:defs}
The digital twin interfaces with an external control system, where the following are the key parameters and in-simulation calculations that are communicated to every controller:

All controllers operate over a shared service-road network modeled as a directed graph $G=(N,B)$, represented in Fig.~\ref{fig:layout}. The list of service tasks is $I$, where each task $i\in I$ occurs at a stand located at node $n_i\in N$ with service duration $t_i^{\mathrm{svc}}$ and time window
$[t_i^{\mathrm{arr}},\,t_i^{\mathrm{dep}}]$.
The fleet of EGSVs is denoted $V$, and the set of charging depots are denoted by $D\subseteq N$.

Vehicles move with velocity $\nu$. The shortest-path distance between tasks $i$ and $j$ is $d(n_i,n_j)$, where $n_i$ indicates the node of task $i$, and then the corresponding travel time between tasks is
\begin{equation}\label{abm:timeDefs}
t(i,j)=\frac{d(n_i,n_j)}{\nu}.   
\end{equation}
Shortest paths are calculated using Dijkstra's algorithm~\cite{dijkstra2022} and precalculated using the in-built library from Matlab~\cite{matlabGraph}.

Vehicle energy is modeled using battery capacity $E^{\max}$ and minimum reserve $E^{\min}$. Energy spent by driving and performing tasks are given by
\begin{equation}\label{abm:energyDefs}
E^{\mathrm{drv}}(i,j)=P^{\mathrm{drv}}\cdot t(i,j), \qquad
E^{\mathrm{tsk}}(i)=P^{\mathrm{tsk}}\cdot t_i^{\mathrm{svc}}.
\end{equation}
Charging at depot $d\in D$ occurs at a fixed rate $P^{\mathrm{C}}$, with the powers specified in Annex~\ref{annex2}.
The location of a vehicle $\ell(v)$ and its current energy $E_v$ can be queried from the DT at each time step. Each vehicle holds an internal task queue whose size is given by $\operatorname{queueSize}(v)$, which represents the number of pending tasks already assigned to vehicle $v$, excluding the task currently being executed. This quantity is updated dynamically whenever a task is assigned, started, or completed in the simulation.

\section{Controllers}\label{sec:controllers}
In this section, we develop the three control strategies for the EGSVs in the DT: First, a reactive dispatching heuristic assigns tasks based solely on each agent's current state, providing a computationally efficient operational baseline. Second, a Model Predictive Control (MPC) approach solves a rolling-horizon Mixed-Integer Linear Program (MILP), offering optimization-based decisions consistent with established industrial practice in other disciplines. Third, a Monte Carlo Tree Search (MCTS) algorithm performs a lookahead search over feasible actions using the updated DT states, acting as a middle ground between heuristics and optimization-based controllers. All three strategies are designed for deployable, closed-loop operation under continuously evolving airport states, rather than on the solving of static combinatorial problems. The interaction between the DT components and the control methods is illustrated in Fig.~\ref{fig:intro}.

\subsection{Heuristics}\label{Heuristics}
The heuristic controller assigns each incoming task $i$ to the vehicle that can begin service the soonest while ensuring a minimum energy level. For each vehicle $v\in V$, its current node is $\ell(v)$, and the travel time to task $i$ is $t(\ell(v),n_i)$, using the distance model defined in Section~\ref{sec:abm:defs}.
The heuristic controller first queries the nearest \emph{available} vehicle with $E_v \ge E^{\min}$, denoted as $v_\mathrm{a}$.  
If $v_\mathrm{a}$ exists, its relative travel time to task $i$ is
\begin{equation}
t^{\mathrm{trav}}_{v_\mathrm{a},i} = t\bigl(\ell(v_\mathrm{a}), n_i\bigr).
\end{equation}
The nearest \emph{busy} vehicle satisfying $E_v \ge E^{\min}$ is denoted $v_\mathrm{b}$. Let $t^{\mathrm{rem}}_{v_\mathrm{b}}$ be its remaining service time. Its availability time at task $i$ is
\begin{equation}
t^{\mathrm{avail}}_{v_\mathrm{b}} = t^{\mathrm{rem}}_{v_\mathrm{b}} + t\bigl(\ell(v_\mathrm{b}), n_i\bigr).
\end{equation}
The assignment rule selects the vehicle minimizing estimated service start time:
\begin{equation}
v^\star = \arg\min_{v \in V:\,E_v \ge E^{\min}} t^{\mathrm{start}}_v(i),
\end{equation}
where
\begin{equation}
t^{\mathrm{start}}_v(i) =
\begin{cases}
t^{\mathrm{trav}}_{v,i}, & v \text{ available},\\[4pt]
t^{\mathrm{rem}}_{v} + t^{\mathrm{trav}}_{v,i}, & v \text{ busy}.
\end{cases}
\end{equation}
Finally, the task $i$ is assigned to $v^\star$ if $\operatorname{queueSize}(v^\star) \le 0$; otherwise, the task is deferred. The simulation implementation is shown in Algorithm~\ref{alg:heur}.
The charging of the vehicles is performed in-simulation in a rule-based fashion, with a vehicle sent to the nearest depot when its energy level drops below $E^{\min}$.
\begin{algorithm}[H]
\caption{\textsc{HeuristicAssignment}$(i, V, E^{\min})$}
\label{alg:heur}
\begin{algorithmic}[1]
\State \textbf{Input:} task $i$, fleet $V$, reserve energy $E^{\min}$
\State $v_\mathrm{a} \gets \Call{getNearest}{V,\ \text{"available"},\ E_v \ge E^{\min}}$
\State $t^{\mathrm{trav}}_{v_\mathrm{a},i} \gets
\begin{cases}
t(\ell(v_\mathrm{a}), n_i), & v_\mathrm{a} \neq \varnothing,\\
\infty, & \text{otherwise.}
\end{cases}$
\State $v_\mathrm{b} \gets \Call{getNearest}{V,\ \text{"busy"},\ E_v \ge E^{\min}}$
\If{$v_\mathrm{b} \neq \varnothing$}
  \State $t^{\mathrm{rem}}_{v_\mathrm{b}} \gets \Call{remainingTime}{v_\mathrm{b}}$
  \State $t^{\mathrm{trav}}_{v_\mathrm{b},i} \gets t(\ell(v_\mathrm{b}), n_i)$
  \State $t^{\mathrm{avail}}_{v_\mathrm{b}} \gets t^{\mathrm{rem}}_{v_\mathrm{b}} + t^{\mathrm{trav}}_{v_\mathrm{b},i}$
\Else
  \State $t^{\mathrm{avail}}_{v_\mathrm{b}} \gets \infty$
\EndIf
\If{$v_\mathrm{b} \neq \varnothing$ \textbf{and} $t^{\mathrm{avail}}_{v_\mathrm{b}} \le t^{\mathrm{trav}}_{v_\mathrm{a},i}$ \textbf{and} $\operatorname{queueSize}(v_\mathrm{b}) \le 0$}
  \State $\Call{assign}{v_\mathrm{b},\ i}$
\ElsIf{$v_\mathrm{a} \neq \varnothing$}
  \State $\Call{assign}{v_\mathrm{a},\ i}$
\Else
  \State $\Call{assign}{v_\mathrm{b},\ i}$ \Comment{\scriptsize{fallback to $v_\mathrm{b}$ even if task queue is not empty}}
\EndIf
\end{algorithmic}
\end{algorithm}

\subsection{Model Predictive Control}\label{sec:mpc}
The MPC controller scheme solves a rolling-horizon mixed-integer linear program (MILP) at each decision epoch. At each step it selects assignments and routes for the EGSV fleet to service tasks, while respecting vehicle energy and time constraints. Given the set up of the DT outlined in Section~\ref{sec:abm:defs} (road network $G$, the set of tasks  \(I\), the set of vehicles \(V\), and the set of depots \(D\)), a set of feasible task–to–task arcs \(\mathcal{E} \subset I \times I\) is defined such that \((i,j)\in \mathcal{E}\) indicates that task \(j\) may be carried out directly after task \(i\).
The controller operates in continuous time. Decision epochs occur at times $t_k = k\,T^{\mathrm{step}}$, $k = 0, 1, 2, \ldots$, where the control step ($T^{\mathrm{step}} > 0$) is the fixed interval between successive re-optimisations. At epoch $t_k$, the planning horizon $[t_k,\, t_k + T^{\mathrm{hor}})$ defines the lookahead window of tasks. The task subset within the horizon for a given time is given by $I_{t} = \bigl\{i \in I \mid t \leq a_i < t + T^{\mathrm{hor}}\bigr\}$, where $a_i$ denotes the arrival time of task $i$. The MILP is solved over $I_{t}$, the resulting assignments are applied, and the horizon then shifts forward to $[t_{k+1},\, t_{k+1}+T^{\mathrm{hor}})$ at the next epoch. The procedure repeats until the simulation end time $t^{\mathrm{end}}$ is reached.
Each task \(i \in I\) occurs at node \(n_i \in N\), has service duration \(t_i^{\mathrm{svc}}\), an availability window \([t_i^{\mathrm{arr}},\,t_i^{\mathrm{dep}}]\), and a reference start time \(t_i^{\mathrm{ref}}\). 
Vehicles serve their assigned tasks sequentially, with their travel times given by~\eqref{abm:timeDefs} and energy consumptions given by~\eqref{abm:energyDefs}.

\subsubsection{Decision variables}
Decision variables include binary \(x^{\mathrm{first}}_{iv}\in\{0,1\}\) indicating whether vehicle \(v\in V\) performs task \(i\) as its first route assignment, and \(x_{ij}\in\{0,1\}\) indicating whether task \(j\) follows task \(i\). The real variable \(t_i^{\mathrm{start}}\) is the start time of task \(i\) and must lie in the window
\[
t_i^{\mathrm{start}}\in[t_i^{\mathrm{arr}},\,t^{\mathrm{end}}],\;\forall i\in I.
\]
Variable \(\xi_i\) denotes the pre-task energy level, non-negative \(e_i\ge0\) is the charge energy added after task \(i\), \(\xi_i^{\mathrm{end}}\) is the energy at end of routing for task \(i\), \(\delta_i\) measures deviation from reference start time, and \(z_{ij}\) is the non-negative idle gap between tasks \(i\) and \(j\). We further introduce a binary variable \(y_i\in\{0,1\}\) indicating whether a charging detour is activated after task \(i\), and non-negative variables \(z^{\mathrm{trav}}_{ij}\) and \(z^{\mathrm{chg}}_{ij}\) denoting the detour–travel time and charging time, respectively, on arc \((i,j)\in\mathcal{E}\).

\subsubsection{Flow conservation constraints}
Vehicle flow conservation is enforced by ensuring each task has exactly one predecessor or is first in a route,
\begin{align}
\sum_{v\in V}x^{\mathrm{first}}_{iv}+\sum_{j:(j,i)\in\mathcal{E}}x_{ji}=1,
&& \forall i\in I,
\label{eq:u_flow_in}\\
\sum_{j:(i,j)\in\mathcal{E}}x_{ij}\le 1,
&& \forall i\in I.
\label{eq:u_flow_out}
\end{align}
Each vehicle may have at most one first assignment,
\begin{align}
\sum_{i\in I}x^{\mathrm{first}}_{iv}\le1,
\qquad \forall v\in V.
\label{eq:u_flow_cap1}
\end{align}
The number of active routes does not exceed fleet size,
\begin{align}\label{eq:u_flow_cap2}
  \sum_{v\in V}\sum_{i\in I}x^{\mathrm{first}}_{iv}\le |V|.
\end{align}
Prior assignments \(\mathcal{L}\) from the previous epoch are fixed:
\begin{equation}
x^{\mathrm{first}}_{iv}=1,
\qquad \forall (i,v)\in\mathcal{L}.
\label{eq:u_lock_first}
\end{equation}

\subsubsection{Time feasibility and schedule deviation constraints}
The start times must satisfy temporal precedence between consecutive tasks. To this end we impose the big-$M$ constraint:
\begin{equation}
t_j^{\mathrm{start}}\ge t_i^{\mathrm{start}}+t_i^{\mathrm{svc}}+t(i,j)-M^{\mathrm{time}}_{ij}(1-x_{ij}),
\qquad \forall (i,j)\in\mathcal{E},
\label{eq:u_time_cons}
\end{equation}
where $M^{\mathrm{time}}_{ij}$ is a big-$M$ constant defined as $M^{\mathrm{time}}_{ij}=\max\{0,\,t_j^{\mathrm{dep}}-(t_i^{\mathrm{arr}}+t_i^{\mathrm{svc}}+t(i,j))\}$. This constraint enforces that if $x_{ij}=1$ (i.e., tasks $i$ and $j$ are assigned consecutively to the same vehicle), then the start time of task $j$ must be at least the end time of task $i$ plus the travel time between them. If $x_{ij}=0$, the constraint becomes non-binding.
The deviation of the optimized start time from the reference schedule is bounded by
\begin{equation}
|t_i^{\mathrm{start}}-t_i^{\mathrm{ref}}| \leq \delta_i,
\qquad \forall i\in I.
\label{eq:u_dev}
\end{equation}
If a task is locked as the first of a vehicle, its start time is fixed:
\begin{equation}
t_i^{\mathrm{start}}=t_i^{\mathrm{ref}},
\qquad \forall i\ \text{such that}\ \exists v:(i,v)\in\mathcal{L}.
\label{eq:u_fix_first_time}
\end{equation}
Constraints \eqref{eq:u_time_cons}–\eqref{eq:u_fix_first_time} preserve chronological feasibility and consistency with the previous solution. Lateness beyond the nominal window end \(t_i^{\mathrm{dep}}-t_i^{\mathrm{svc}}\) is allowed
but penalized through \(\delta_i\) in the objective.

\subsubsection{Energy constraints}
For each vehicle $v\in V$, the initial energy is $E_v^{0}$ and its previous node is $n_v^{\mathrm{old}}$. The energy required to reach its current task $i$ from its previous location is
\begin{equation}
E^{\mathrm S,\mathrm{init}}(v,i)=P^{\mathrm{drv}}\cdot t(n_v^{\mathrm{old}},n_i).
\end{equation}
We define an auxiliary variable $\xi_i$ that records the energy of a vehicle before executing task $i$. This variable will be used to ensure the tours proposed for each vehicle are energy-feasible. We ensure all vehicles never exceed the operational limits by imposing $\xi_i\in[E^{\min},E^{\max}]$.
For tasks that come first in a route, the pre-task energy equals the initial energy of the assigned vehicle minus the travel energy from its previous position:
\begin{align}
\xi_i &\le \sum_{v\in V} x^{\mathrm{first}}_{iv}\bigl(E_v^{0}-E^{\mathrm S,\mathrm{init}}(v,i)\bigr)
    +E^{\max}\Bigl(1-\sum_{v\in V}x^{\mathrm{first}}_{iv}\Bigr), \nonumber\\
\xi_i &\ge \sum_{v\in V} x^{\mathrm{first}}_{iv}\bigl(E_v^{0}-E^{\mathrm S,\mathrm{init}}(v,i)\bigr)
    -E^{\max}\Bigl(1-\sum_{v\in V}x^{\mathrm{first}}_{iv}\Bigr),\nonumber\\
& \qquad \qquad \qquad \qquad \qquad \qquad \qquad \qquad \qquad  \forall i\in I.
\label{eq:u_first_soc}
\end{align}

We define the minimal detour time and energy between tasks through the nearest depot as
\begin{equation}\label{eq:u_ener_depot}
t^{\mathrm C}(i,j)=\min_{d\in D}\bigl\{t(i,d)+t(d,j)\bigr\},\quad
E^{\mathrm C}(i,j)=P^{\mathrm{drv}}\,t^{\mathrm C}(i,j).
\end{equation}
Detour travel time is enforced when both the arc \((i,j)\) is chosen and
the detour binary \(y_i\) is activated.
\begin{align}
z^{\mathrm{trav}}_{ij}
&\ge t^{\mathrm C}(i,j)\,y_i - M^{\mathrm{time}}_{ij}(1-x_{ij}),
\qquad \forall (i,j)\in\mathcal{E}, \label{eq:detour_time_lb}\\[1ex]
z^{\mathrm{trav}}_{ij}
&\le t^{\mathrm C}(i,j)\,y_i + M^{\mathrm{time}}_{ij}(1-x_{ij}),
\qquad \forall (i,j)\in\mathcal{E}. \label{eq:detour_time_ub}
\end{align}
To ensure that charging implies a detour, we impose
\begin{equation}
e_i \;\le\; E^{\max} y_i,
\qquad \forall i\in I.
\label{eq:y_charge_link}
\end{equation}
Defining $E^{\mr{det}}_{i,j} = P^{\mathrm{drv}}\,z^{\mathrm{trav}}_{ij}$, the energy continuity for a successor tasks now becomes
\begin{align}
\xi_j &\le \xi_i - E^{\mathrm{tsk}}_i
      - E^{\mathrm{drv}}_{i,j}
      - E^{\mathrm{det}}_{i,j}
      + e_i
      + E^{\max}(1-x_{ij}), \nonumber\\
\xi_j &\ge \xi_i - E^{\mathrm{tsk}}_i
      - E^{\mathrm{drv}}_{i,j}
      - E^{\mathrm{det}}_{i,j}
      + e_i
      - E^{\max}(1-x_{ij}), \nonumber\\
&\qquad\qquad\qquad\qquad\qquad\qquad\qquad\qquad\forall (i,j)\in\mathcal{E}.
\label{eq:u_soc_seq}
\end{align}
Equations \eqref{eq:u_first_soc}–\eqref{eq:u_soc_seq} ensure energy consistency for both initial and consecutive tasks, while accounting for the energy used in any detour and allowing continuous charging decisions after tasks.
Additionally, at the end of the route the vehicle’s energy must be non-negative: let $y_i^{\mathrm{succ}}=\sum_{j:(i,j)\in\mathcal{E}}x_{ij}$ indicate whether task $i$ has a successor. Then the end-of-route energy must satisfy
\begin{align}
\xi_i^{\mathrm{end}}&\le E^{\max}\bigl(1-y_i^{\mathrm{succ}}\bigr),\qquad \forall i\in I,\\
\xi_i^{\mathrm{end}}&\ge \xi_i-E^{\max}y_i^{\mathrm{succ}},\qquad \forall i\in I.
\label{eq:u_end_soc}
\end{align}

\subsubsection{Charging constraints}
With $z_{ij}$ as the idle time (if available) between tasks $i$ and $j$, it satisfies
\begin{align}
&z_{ij}\ge t_j^{\mathrm{start}}-t_i^{\mathrm{start}}-t_i^{\mathrm{svc}}-t(i,j),
\quad \forall (i,j)\in\mathcal{E},\\
&z_{ij}\le t_j^{\mathrm{start}}-t_i^{\mathrm{start}}-t_i^{\mathrm{svc}}-t(i,j)
  +M^{\mathrm{time}}_{ij}(1-x_{ij})\\ 
  & \qquad\qquad\qquad\qquad\qquad\qquad\qquad\qquad \forall (i,j)\in\mathcal{E},\nonumber\\
&z_{ij}\ge0 \quad \forall (i,j)\in\mathcal{E}.
\label{eq:u_zgap}
\end{align}
This slack is decomposed as
\begin{equation}
z_{ij} = z^{\mathrm{trav}}_{ij} + z^{\mathrm{chg}}_{ij},
\qquad z^{\mathrm{trav}}_{ij}\ge 0,\;
z^{\mathrm{chg}}_{ij}\ge 0,
\quad \forall (i,j)\in\mathcal{E}.
\label{eq:z_decomp}
\end{equation}

Charging time is only available when detour mode is active. We enforce this with a simple linking constraint
\begin{equation}
\sum_{j:(i,j)\in\mathcal{E}} z^{\mathrm{chg}}_{ij} \;\le\; M^{\mathrm{time}}\, y_i,
\qquad \forall i\in I,
\label{eq:chg_requires_detour}
\end{equation}
where $M^{\mathrm{time}}$ is a suitably large time constant (e.g., the horizon length).

The charge energy $e_i$ is then limited by the usable positive charging time through
\begin{equation}
e_i\le 
P^{\mathrm C}
\sum_{j:(i,j)\in\mathcal{E}} z^{\mathrm{chg}}_{ij}
- M^{\mathrm{ener}} (1-y_i),
\qquad \forall i\in I,
\label{eq:u_charge_gap}
\end{equation}
where $M^{\mathrm{ener}}$ is a big-M constant defined as $M^{\mathrm{ener}}=E^{\mathrm{max}}$. This constraint ensures that if $y_i=0$ no detour and thus no charging occurs. If $y_i=1$, then $e_i$ is limited by the available charging time multiplied by the charger power. Together, \eqref{eq:u_zgap}–\eqref{eq:u_charge_gap} guarantee that charging energy is only drawn when positive idle time exists, that it is limited by charger power and the available slack between consecutive tasks, and that any charging implies activation of a detour with sufficient travel and charging time (via \eqref{eq:y_charge_link} and \eqref{eq:chg_requires_detour}).

\subsubsection{Objective}
The objective function comprises three weighted terms. The first term minimizes the deviation $\delta_i$ from each task’s previous start time, promoting consistency across successive assignments. 
The second term minimizes the recharged energy $e_i$, reducing total energy consumption and charging events. 
The third term maximizes the end-of-route energy $\xi_i^{\mathrm{end}}$, ensuring adequate battery level at the horizon’s end. With this we write our final objective as follows:
\begin{equation}\label{eq:u_obj}
J=\;
w_1\sum_{i\in I}\delta_i
+w_2\sum_{i\in I}e_i
-w_3\sum_{i\in I}\xi_i^{\mathrm{end}}.
\end{equation}

\subsubsection{Final MILP formulation}
Combining all the above, we summarize the optimization problem as follows:

\textbf{Problem~1}~(ESGV Routing with Charging)  
\textit{Given a set of vehicles $V$, refuelling tasks $I$, precomputed feasible arcs $\mathcal{E}\subseteq I\times I$, travel times $t(\cdot,\cdot)$, energy parameters $(E^{\max},P^{\mathrm{drv}},P^{\mathrm{tsk}},P^{\mathrm C})$, the routing, timing, and charging decisions 
minimizing the objective $J$ result from}
\begin{equation}
\begin{aligned}
\label{eq:final-unified}
\min_{\substack{
\,x_{ij},\,\xi_i,\,y_i,\\\,e_i,\,\xi_i^{\mathrm{end}},\,\delta_i
}}
\quad  
& J(w_1,w_2,w_3)\\
\textrm{s.t.}\quad 
& \eqref{eq:u_flow_in}\text{--}\eqref{eq:u_fix_first_time},\;
  \eqref{eq:u_first_soc}\text{--}\eqref{eq:u_soc_seq},\;
  \eqref{eq:u_zgap}\text{--}\eqref{eq:u_end_soc},\\[1mm]
& x^{\mathrm{first}}_{iv}\in\{0,1\} \quad \forall i\in I,\ \forall v\in V,\\
& x_{ij}\in\{0,1\} \quad \forall (i,j)\in\mathcal{E},\\
& y_i\in\{0,1\} \quad \forall i\in I,\\
& e_i\ge 0,\;\; z_{ij}\ge 0,\quad \forall (i,j)\in\mathcal{E},\\
& \xi_i^{\mathrm{end}}\ge 0,\;\; \delta_i\ge 0
\quad \forall i\in I.
\nonumber
\end{aligned}
\end{equation}
This MILP can be solved with off-the-shelf optimizers and is re-solved at each decision time $kT^{\mathrm{step}}$ with the set $I_t=\{i\in I \mid t_i^{\mathrm{arr}}\in[kT^{\mathrm{step}},\,kT^{\mathrm{step}}+T^{\mathrm{hor}}]\}$.

\subsubsection{Computational Tractability}
To guarantee running times to be suitable for online implementation, the arc set
$\mathcal{E}$ is constructed by a spatial, temporal and energy filtering based on the problem setup.
Each task $i$ is associated with a pier identifier $p_i$ (e.g., gate G09 corresponds to pier G). For each task $i$ we construct a candidate successor list by sorting feasible tasks $j$ by pier compatibility (tasks with $p_i=p_j$ first) and by slack time between tasks $i$ and $j$ defined as
\begin{equation}
t^{\mathrm{slack}}_{ij}
= t_j^{\mathrm{arr}} - \bigl(t_i^{\mathrm{arr}} + t_i^{\mathrm{svc}} + t(i,j)\bigr).
\end{equation}
Task pairs with very small slacks are discarded as temporally infeasible. Similarly, task pairs where the energy required to travel from $i$ to $j$ exceeds the vehicle's $E_v$ (minus a safety margin) are also discarded.
From this list we retain only the first $K$ successors, ordered by pier compatibility, slack, and travel time. The resulting filtered arc set $\mathcal{E}$ guarantees computational tractability at runtime.
\subsection{Monte Carlo Tree Search (MCTS)}\label{sec:mcts}

The controller explores future routing and charging decisions by performing a tree-search over the set of feasible states. We cast the problem as a Markov decision process \((\mathcal{S},\mathcal{A},R)\), where \(\mathcal{S}\) is the set of feasible states, \(\mathcal{A}\) the set of admissible actions \(a=(v,i)\) representing either a vehicle–task assignment or a charging move, and \(R\colon\mathcal{S}\times\mathcal{A}\to[R^{\min},R^{\max}]\) the reward function, with \(R^{\min}\) and \(R^{\max}\) denoting the minimum and maximum per-step reward, respectively.

A state \(s\in\mathcal{S}\) encodes the current time \(t\), the set of tasks within the lookahead window \(T^{\mathrm{hor}}\): $I^{\mathrm{hor}} = \bigl\{i\in I \mid t_i^{\mathrm{arr}}\in[t,\;t+T^{\mathrm{hor}})\bigr\}$, each vehicle’s location and energy level, and the partial routes already committed. 
The search tree is built at every control step \(T^{\mathrm{step}}_{\mathrm{MCTS}}\), spanning the planning horizon \([t,\;t+T^{\mathrm{hor}})\), where state transitions within the simulation are advanced in increments of \(T^{\mathrm{SIM}}_{\mathrm{MCTS}}\), the temporal resolution used to step between successive states during tree rollout.
In the following we describe the main phases of the MCTS algorithm, followed by the action-generation mechanism and the reward structure. 
\begin{figure}%
    \centering
    \includegraphics[width=\linewidth]{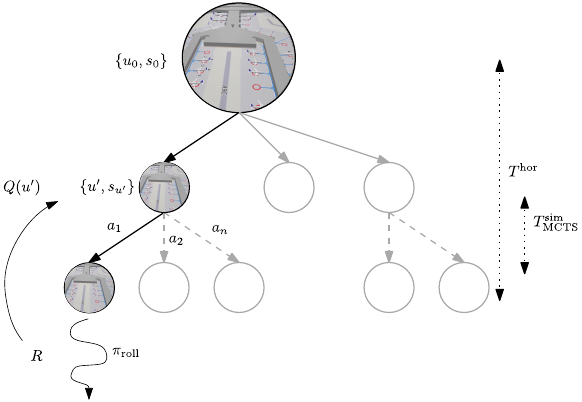}
    \caption{Overview of the MCTS search procedure. Starting from the root node \(u_0\) with the airport DT state \(s_0\), the algorithm iteratively selects child nodes using the UCT policy (Selection), expands unvisited nodes by sampling feasible actions (Expansion), simulates complete schedules using a rollout policy (Simulation), and backpropagates the obtained rewards $R$ up the tree (Backpropagation). After exhausting the computational budget, the best action at the root is executed in the real system. The procedure is then repeated from the advanced DT state.}
    \label{fig:sim}  
\end{figure}
\subsubsection{MCTS Algorithm}
Starting from the root node \(u_0\) with state \(s_0\) (no tasks yet dispatched), the algorithm iterates the following four phases until the computational budget is exhausted. The pseudocode is presented in Algorithm~\ref{alg:mcts} and illustrated in Fig. \ref{fig:sim}. The implementation follows~\cite{kocsis-2006-bandit}.

\textit{Selection:} From a non-terminal node $u$ with state $s$, a child node $u'$ is selected using the Upper Confidence bounds applied to Trees (UCT) policy,
\begin{equation}
\text{UCT}(u') = \arg\max_{u' \in \text{children}(u)} \left( \frac{Q(u')}{C(u')} + \gamma^{\mr{expl}} \sqrt{ \frac{\ln C(u)}{C(u')} } \right)
\label{eq:uct}
\end{equation}
where $C(u')$ and $C(u)$ count the number of visits to node $u'$ and its parent node $u$, respectively, $Q(u')$ accumulates the total reward obtained through node $u'$ ($Q(u') = \sum_{i=1}^{C(u')} R^{(i)}$), and $\gamma^{\mr{expl}}$ is the exploration constant that controls the trade-off between exploitation and exploration.
Starting at the root node $u_0$, the algorithm iteratively selects the child node with the highest UCT value via~\eqref{eq:uct} until reaching an unexpanded node (with unvisited actions) or a terminal node. 

\textit{Expansion:} When an unexpanded node is reached, a feasible action is sampled to create a new child node. An untried action $a\in\mathcal{A}(s_u)$ is selected according to a given expansion policy, where the procedure to generate feasible actions considers vehicle availability, task urgency, and energy constraints, detailed in Section~\ref{mcts:agen}. Then, the transition to the resulting successor state is obtained by applying the action and stepping the DT forward by $T_{\mathrm{MCTS}}^{\mathrm{SIM}}$.

\textit{Simulation:} From the expansion node $s_u$, a complete schedule is generated and simulated, using a rollout policy $\pi^{\mathrm{roll}}$. This policy is executed until a terminal state is reached or the time horizon $T^{\mr{hor}}$ is met, accumulating the discounted return from \eqref{eq:return}. The reward $R$ is computed as
\begin{equation}
R \;=\; \sum_{k=0}^{T^{\mr{hor}}-1} \gamma^{\mr{dis}}_{\,k}\, r_{t+k+1},
\label{eq:return}
\end{equation}
where the rollout terminates at horizon $T^{\mr{hor}}$ or a terminal state and $0<\gamma^{\mr{dis}}\le 1$ is the discount factor. The immediate reward $r_{t+k+1}$ is defined in Section~\ref{mcts:reward}.

\textit{Backpropagation:} The cumulative reward obtained in simulation is propagated up the tree, updating $Q(u)$ and $C(u)$ for every node on the selection path. For each node $u$ on the path from $u'$ up to $u_0$ we do
\begin{equation}
C(u) \leftarrow C(u)+1,\qquad Q(u) \leftarrow Q(u)+R.
\end{equation}
Finally, the best action at the root (the one with the highest empirical mean return $Q(u')/C(u')$) is executed in the real system, the horizon tasks $I_t$ are updated, and MCTS is re‐invoked at the next decision step. 
\begin{algorithm}
\caption{MCTS Scheduler at step $t$}\label{alg:mcts}
\begin{algorithmic}[1]
\Require state $s_t$, computational budget $T^{\mathrm{budget}}$
\State initialize root node $u_0$ with state $s_{u_0}\gets s_t$, and $C(\cdot)\gets 0$, $Q(\cdot)\gets 0$
\While{time $< T^{\mathrm{budget}}$}
    \State $P \gets \textsc{Select}(u_0)$ \Comment{using UCT from~\eqref{eq:uct}}
    \State $u \gets \textsc{last}(P)$
    \If{$u$ is unexpanded and nonterminal}
        \State $a \gets \textsc{Expand}(s_u)$ \Comment{$a\in\mathcal{A}$ from Alg.~\ref{alg:agen}}
        \State ($u',s_{u'}$) $\gets \textsc{SimStep}(s_u,a)$
    \Else
        \State $u' \gets u$
    \EndIf
    \State $R \gets \textsc{Simulate}(u')$ \Comment{Reward as in~\eqref{eq:return}}
    \State $\textsc{Backpropagate}(u', R)$
\EndWhile
\State \Return $I_t^\star \in \argmax_{a\in \mathcal{A}(s_t)} \frac{Q(u_a)}{C(u_a)}$
\end{algorithmic}
\end{algorithm}

\subsubsection{Action Generation}\label{mcts:agen}
The search tree depends on how feasible actions are generated for each state. At decision time \(t\), an action assigns a vehicle \(v\in V\) either to a task \(i\in I_t\) or sends \(v\) to charge at a depot \(d\in D\). To control the branching factor, the action generator returns at most \(N_{\max}\) candidate actions ranked by a combined scoring function.

\textit{Serviceable tasks:} A task \(i\in I_t\) is considered serviceable if \(t_i^{\mathrm{arr}}\le t\) and its demand \(d_i>0\). When no such tasks exist, tasks arriving within a look-ahead window \(H\) are included to enable preemptive dispatching.

\textit{Feasible action set generation:}
For vehicle–task pair \((v,i)\), compute travel time \(t(\ell(v),n_i)\). The pair is feasible if vehicle energy \(E_v\) suffices to perform service and reach the nearest charger.
Greedy matching assigns each flight to the idle vehicle with minimal travel time $t\bigl(\ell(v),n_i\bigr)$, where flights are ranked by urgency $(t - t^{\mr{dep}}_i)$.
In addition, multi-dispatch actions can be formed by combining two or more single dispatches, combining up to $M_{\max}$ single dispatches.
Vehicles with $E_v \leq E^{\min}$ can also charge, and idle vehicles can return to the depot when the candidate set is empty.

\textit{Action filter:} The action generator limits the total number of actions returned to the MCTS procedure to a maximum $N_{\max}$, keeping only those with the best heuristic scores $\operatorname{Score}(a)$ that combine travel time, urgency, and energy considerations. Each action $a$ receives a score
\begin{equation}
\Score(a) = w_t\,\phi^{\mathrm{t}}(a) + w_u\,\phi^{\mathrm{u}}(a) + w_e\,\phi^{\mathrm{e}}(a),
\end{equation}
where \(\phi^{\mathrm{t}}\) is the travel-time cost, \(\phi^{\mathrm{u}}\) the urgency term, and \(\phi^{\mathrm{e}}\) the energy term. Only the top-ranked \(N_{\max}\) actions are passed to the expansion phase of MCTS.
The complete action generation procedure is summarized in Algorithm~\ref{alg:agen}.

\begin{algorithm}
\caption{\textsc{Action Generator}}
\label{alg:agen}
\begin{algorithmic}[1]
\Require state $s$ at time $t$, parameters $(H,T^{\mr{hor}},M_{\max},C_{\max},N_{\max})$
\State $\bar{I} \gets \AssignedTasks(s)$          
\State $\bar{V} \gets \LockedVehicles(s)$       
\State $I_t^{\mr{elig}} \gets \EligibleDepartures(I \setminus \bar{I},H,T^{\mr{hor}})$
\State $\mathcal{A}^{\mathrm{disp}} \gets \GreedyDispatches(s,I_t^{\mathrm{elig}},\bar{I},\bar{V})$
\State $\mathcal{A}^{\mathrm{chg}} \gets \{ (v,d) \mid v\in V,\ d\in D,\ E_v \le E_{\min}\}$
\State $\mathcal{A}^{\mathrm{depot}} \gets \{ (v,d_{depot})\mid v\in V\setminus \bar{V}\}$
\State $\mathcal{A}^{\mathrm{multi}} \gets \MultiDispatch(\mathcal{A}^{\mathrm{disp}}, M_{\max}, C_{\max})$
\State $\mathcal{A} \gets \text{Idle} \cup \mathcal{A}^{\mathrm{chg}} \cup \mathcal{A}^{\mathrm{disp}} \cup \mathcal{A}^{\mathrm{multi}} \cup \mathcal{A}^{\mathrm{depot}}$
\ForAll{$a \in \mathcal{A}$}
    \State $S(a) \gets \Score(a)$
\EndFor
\State \Return $\argmax\limits_{\substack{\mathcal{B}\subseteq \mathcal{A}\\|\mathcal{B}|=N_{\max}}} \sum_{a\in \mathcal{B}} S(a)$ \Comment{top $N_{\max}$ actions}
\end{algorithmic}
\end{algorithm}

\subsubsection{Reward Formulation}\label{mcts:reward}
Given current state $s$, action $a$, and successor state $s'$, the immediate reward $R'(s,a,s')$ is given by the sum of a completion reward, a post-departure penalty, a travel penalty, and a potential-based shaping term.

\textit{Completion reward:} For every task $i\in I$ that becomes fully serviced in this step (i.e., serviced in $s'$ but not in $s$), we give an early/on-time reward with exponential decay in the last window before departure
\begin{equation}
R^{\mathrm{comp}} = \sum_{i \in I} r^{\mathrm{comp}}_i,
\end{equation}
where
\begin{equation}
r^{\mathrm{comp}}_i =
\begin{cases}
R_{\max}, & t_i^{\mathrm{comp}} \le t_i^{\mathrm{dep}} - t^{\mr{early}} ,\\[2pt]
R_{\max} \cdot \frac{\exp(\lambda \alpha_i)-1}{\exp(\lambda)-1}, & t_i^{\mathrm{dep}} - t^{\mr{early}} < t_i^{\mathrm{comp}} \le t_i^{\mathrm{dep}},\\[4pt]
0, & \text{otherwise,}
\end{cases}
\end{equation}
with $R_{\max}$ as a maximum reward for this component, $t^{\mr{early}}$ is a safety margin, $\lambda$ is an exponential decay rate, and
\begin{equation}
\alpha_i = \frac{t_i^{\mathrm{dep}} - t_i^{\mathrm{comp}}}{t^{\mr{early}}} \in [0,1].
\end{equation}

\textit{Post-departure penalty:} For tasks that are still unserved after their departure time we apply an exponential penalty
\begin{equation}
R^{\mathrm{post}} = \sum_{i \in I} \psi_i,
\end{equation}
with
\begin{equation}
\psi_i =
\begin{cases}
0, & \text{if } t' \le t_i^{\mathrm{dep}} \text{ or } i\in{\bar{I}},\\[2pt]
-w^{\mathrm{post}} \cdot \bigl(e^{\lambda^{\mathrm{post}} (t' - t_i^{\mathrm{dep}})} - 1\bigr), & \text{otherwise},
\end{cases}
\end{equation}
where $w^{\mathrm{post}}$ is a scaling factor, $\lambda^{\mathrm{post}}$ is a decay rate, $t'$ is the time in $s'$. A maximum cap is also applied.

\textit{Travel penalty:} We penalize total travel minutes of all vehicles between $s$ and $s'$:
\begin{equation}
R^{\mathrm{trav}} = - w^{\mathrm{trav}} \cdot T^{\mathrm{trav}},
\end{equation}
where $w^{\mathrm{trav}}$ is a weighting factor and $T^{\mathrm{trav}}$ is the total travel time of all vehicles in minutes.
With this our immediate reward is given by
\begin{equation}
R'(s,a,s') = R^{\mathrm{comp}} + R^{\mathrm{post}} + R^{\mathrm{trav}}.
\end{equation}
To this we add a potential-based term~\cite{ng1999} to better guide the search through sparse rewards. The shaped reward is given by:
\begin{equation}
R'(s,a,s') = R(s,a,s') + \gamma^{\mr{shp}}\bigl(\phi^{\mr{s}}(s') - \phi^{\mr{s}}(s)\bigr),
\end{equation}
where $\phi^{\mr{s}}(s)$ depends on the total unserved flights and the total vehicle energy in a given state $s$. In particular, we consider two potential-based shaping terms:

\textit{Outstanding-demand shaping:} Let $F(s)$ be the total unserved flights in $s$. Then we define $R^{\mathrm{out}}$ as
\begin{equation}
R^{\mathrm{out}} = w_{\mathrm{out}}\bigl(F(s) - F(s')\bigr),
\end{equation}
with $w_{\mathrm{out}}$ a weighting factor.

\textit{Fleet Energy shaping:} Let $E^{\mathrm{T}}(s)$ be the total energy across all vehicles in state $s$. Then we define $R^{\mathrm{E}}$
\begin{equation}
R^{\mathrm{E}} = w_{\mathrm{E}}\bigl(E^{\mathrm{T}}(s') - E^{\mathrm{T}}(s)\bigr),
\end{equation}
with $w_{\mathrm{E}}$ a weighting factor.

The implementation of Algorithm~\ref{alg:mcts} using the reward structure from Section~\ref{mcts:reward} and the action generation mechanism from Algorithm~\ref{alg:agen} results in the MCTS controller for the ESGV scheduling and charging problem. 
This implementation has a controlled complexity (branching factor is bounded by $N_{\max}$), making it suitable for real-time decision-making.

\section{Results}\label{sec:Results}
We present the results from controller tests on the digital twin focusing on the fleet of Hydrant-dispenser EGSVs.
Experiments vary in schedule scenarios and fleet sizes. The primary outputs are flight delays, driving energy consumption and charging metrics.
\subsection{Experimental Setup}
We tested three representative days: 02-03-2024, 10-03-2024, and 07-11-2024, with 355, 387 and 404 aircraft handled, respectively, which were found to be representative of low, average and busy airport operations.
Table~\ref{tab:schedule_statistics} summarizes the key statistics of the flight schedules. 
\begin{table}[t]
  \centering
  \small
  \caption{Flight schedule statistics}
  \label{tab:schedule_statistics}
  \setlength{\tabcolsep}{3pt}
  \begin{threeparttable}
    \begin{tabular*}{\linewidth}{@{\extracolsep{\fill}}llllll@{}}
      \toprule
      \begin{tabular}[c]{@{}l@{}}Schedule\\Date\end{tabular} &
      \begin{tabular}[c]{@{}l@{}}No. of\\Aircraft\end{tabular} &
      \begin{tabular}[c]{@{}l@{}}Total\\Flights\end{tabular} &
      \begin{tabular}[c]{@{}l@{}}Total Time\\Changes\end{tabular} &
      \begin{tabular}[c]{@{}l@{}}Total Gate\\Changes\end{tabular} &
      \begin{tabular}[c]{@{}l@{}}Avg. Hourly\\Departures\end{tabular} \\
      \midrule
      02-03-2024 & 355 & 1103 & 469 & 532 & 26.29 \\
      10-03-2024 & 387 & 1224 & 566 & 672 & 29.14 \\
      07-11-2024 & 404 & 1294 & 875 & 910 & 27.04 \\
      \bottomrule
    \end{tabular*}
    \begin{tablenotes}[flushleft]
      \footnotesize
      \item Note: Time changes reflect only deviations in departure/arrival times higher than 10 min from schedule. Source: Schiphol API~\cite{SchipholAPIs}.
    \end{tablenotes}
  \end{threeparttable}
\end{table}
Fleets of 20, 25, and 30 ESGVs were evaluated (20 chosen as the smallest fleet that the heuristic could handle on the busiest day).
For each fleet size tested we allow an equal number of chargers, reflecting airport practice and ensuring no capacity constraints are imposed by charger availability.
Controllers re-solve every 5 min ($T^{\mathrm{step}} = 5$ min), in line with the API refresh rate, with a 120-min prediction horizon ($T^{\mathrm{hor}} = 120$ min) shared by both predictive controllers, and a 5-min compute budget ($T^{\mathrm{budget}}_{\mathrm{MPC}} = T^{\mathrm{budget}}_{\mathrm{MCTS}} = 5$ min).
The heuristic controller has a negligible computation time per re-solve (less than 10 ms.) due to its low complexity and is event-based triggered (only with new aircraft arrivals). 
The MILP problem from the MPC controller is solved using Gurobi 10.1~\cite{gurobi}. A varying optimality gap ranging between 20\% and 0\% was obtained for different re-solves due to the time budget imposed and variation in task demands.
The MCTS controller was implemented by adapting the formulation in~\cite{mctreesearch4j}. For the $T^{\mathrm{budget}}$ and $T^{\mathrm{hor}}$ it resulted in an average of 10,000 Alg.~\ref{alg:mcts} iterations per simulation minute.
All simulations were run on a single server with 16 GB of RAM and an Intel i7-9750H processor (20 cores).
Table~\ref{tab:all-parameters} presents other relevant parameter values used in the DT, MPC and MCTS controllers.
\begin{table}[t]
  \centering
  \small
  \caption{Parameter values for the experiments.}
  \label{tab:all-parameters}
  \setlength{\tabcolsep}{4pt}
  \begin{threeparttable}
    \begin{tabular*}{\linewidth}{@{\extracolsep{\fill}}l c l l c l@{}}
      \toprule
      Symbol & Value & Unit 
        & Symbol & Value & Unit \\
      \midrule
      \multicolumn{6}{l}{\textbf{DT parameters}} \\
      \midrule
      $m_{v}$        & 7,500              & kg  
        & $E_{\max}$      & 90                  & kWh \\
      $E_{\min}$     & 9                   & kWh  
        & $P_{C}$         & 40                  & kW \\
      $q$            & 150                 & m$^3$/h  
        & $p$             & $3.10\times10^{5}$  & Pa \\
      $\nu_{\mathrm{tgt}}$ & 25            & km/h  
        & $a_{x,\max}$    & 1.0                 & m/s$^2$ \\
      $\rho$         & 1.225               & kg/m$^3$  
        & $C_{d}$         & 0.5                 & – \\
      $A$            & 2.5                 & m$^2$  
        & $\mu$           & 0.01                & – \\
      $\eta$         & 0.8                 & –  
        &                 &                     &  \\
      \midrule
      \multicolumn{6}{l}{\textbf{MPC parameters}} \\
      \midrule
      $T^{\mathrm{hor}}$   & 120                & min  
        & $T^{\mathrm{step}}_{\mathrm{MPC}}$  & 5                   & min \\
      $w_{1}$        & 10                  & –  
        & $w_{2}$         & 5                   & – \\
      $w_{3}$        & 1                   & –  
        & $T^{\mathrm{budget}}_{\mathrm{MPC}}$ & 5          & min \\
      \midrule
      \multicolumn{6}{l}{\textbf{MCTS parameters}} \\
      \midrule
      $T^{\mathrm{hor}}$   & 120                & min  
        & $T^{\mathrm{step}}_{\mathrm{MCTS}}$ & 5            & min \\
      $T^{\mathrm{budget}}_{\mathrm{MCTS}}$ & 5       & min  
        & $H$            & 120                 & min \\
      $\gamma^{\mathrm{expl}}$ & 250              & –  
        & $\gamma^{\mathrm{dis}}$ & 0.998         & – \\
      $K$            & 3                   & –  
        & $M_{\max}$     & 4                   & – \\
      $C_{\max}$     & 4                   & –  
        & $N_{\max}$     & 10                  & – \\
      $R^{\min}$     & -1500               & –  
        & $R^{\max}$     & 500                 & – \\
      $t^{\mathrm{early}}$ & 20            & min  
        & $\lambda$      & 3.0                 & – \\
      $w^{\mathrm{post}}$ & 100             & –  
        & $\lambda^{\mathrm{post}}$ & 0.15      & – \\
      $w^{\mathrm{trav}}$ & 2.5             & –  
        & $w_{\mathrm{out}}$ & 20              & – \\
      $w_{\mathrm{E}}$ & 50                 & –  
        &                 &                     &  \\
      \bottomrule
    \end{tabular*}
    \begin{tablenotes}[flushleft]
      \footnotesize
      \item Note: Fleet characteristics refer to ESGVs. Controller settings list the MPC and MCTS configurations used in the experiments. Sources and further details on the parameters can be found in Annex~\ref{annex2}.
    \end{tablenotes}
  \end{threeparttable}
\end{table}

\subsection{Controller Performance}
In this subsection we analyze the controller performance in terms of convergence, flight delays, energy consumption, and the relation between these through charging patterns.

\subsubsection{Convergence}

\begin{figure}[htpb]
\centering
\includegraphics[width=\linewidth]{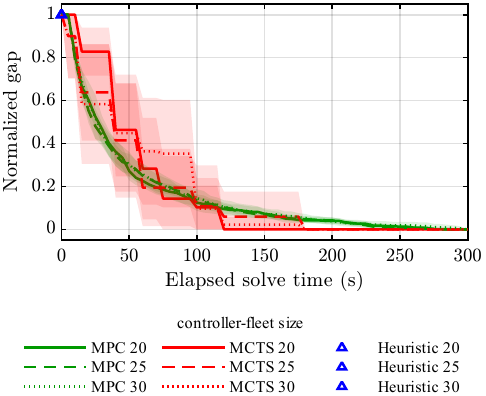}
\caption{Convergence of the MPC and MCTS controllers over iterations for the 300s computational budget, each normalized to its best found solution. MPC improves steadily across iterations, while MCTS converges rapidly in early iterations before plateauing. The heuristic controller is non-iterative and thus has no convergence curve.}
\label{fig:convergence}
\end{figure}

Fig.~\ref{fig:convergence} compares the convergence behaviour of the MPC and MCTS controllers under a 300 s solve budget. MCTS exhibits stepwise convergence, with most improvements occurring in the early part of the budget and final performance reached by approximately 200 s. This suggests that shorter solve budgets may be sufficient for MCTS in practical deployments, although budgets below 100 s would truncate a large part of its improvement. MPC improves more smoothly and consistently, converging faster towards its own best found solution under short solve times, but it does not fully plateau within the 300 s budget. This indicates that MPC can benefit from longer solve times when operational conditions allow, while still remaining competitive under tighter budgets. MCTS also shows substantially higher variability across replications, consistent with the stochastic nature of the algorithm. Across the three fleet sizes, both controllers display similar convergence patterns, indicating stable scalability over the tested range.

\subsubsection{Delays}
\begin{figure}[t]
    \centering
    \includegraphics[width=1\linewidth]{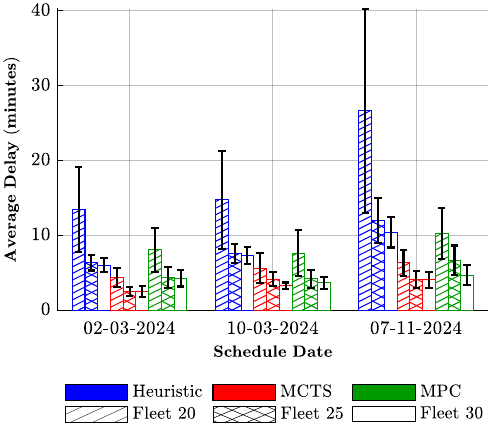}
    \caption{Average flight delays for each controller, for each flight schedule. Error bars show the standard deviation of the delay across all flights of the simulated day. The MCTS controller consistently outperforms both the heuristic and MPC controllers across all scenarios, and also produces the narrowest delay distribution.}
    \label{fig:mpcsched}
\end{figure}

We define the average flight delay of a scenario as the mean departure delay over all flights handled on that day. In terms of delays, Fig.~\ref{fig:mpcsched} shows the average flight delays for each controller across the three schedules and three fleet sizes tested; the error bars report the standard deviation of the delay across all flights of that day.
The delays increase with higher demand levels for all controllers, with a significant jump in the most demanding schedule (07-11-2024). Delays decrease when increasing the fleet from 20 to 25 vehicles, while the improvement from 25 to 30 vehicles is narrow.
The best performance is shown by MCTS in all scenarios, which has a pooled average delay of 4.09 min across all scenarios, an improvement of 7.5 min (64.7\%) and 1.9 min (31.5\%) over the heuristic and MPC controllers, respectively.
The performance of the MPC and MCTS controllers become more similar as the number of vehicles in the fleet increases.
The pooled standard deviation of the flight-delay distribution across all scenarios is 7.83 min for the heuristic, compared to 3.80 min and 2.20 min for the MPC and MCTS controllers, respectively. The heuristic shows large variance under constrained fleets, meaning that its higher average delay is accompanied by a substantially longer tail of badly delayed flights.
\subsubsection{Energy Consumption}
\begin{figure}[t]
    \centering
    \includegraphics[width=1\linewidth]{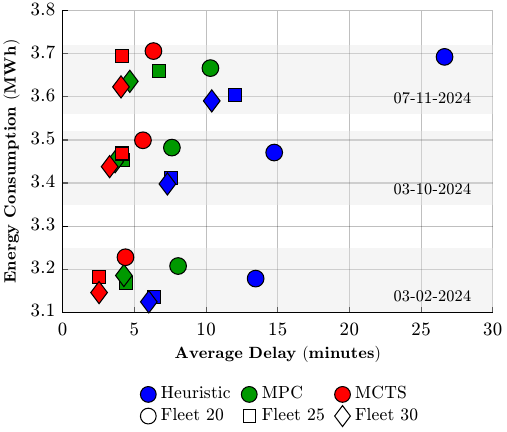}
    \caption{Pareto plot of total energy consumption vs. average delay, for each controller, fleet size and flight schedule (delimited by grey background). MPC and MCTS controller consistently outperform the heuristic in terms of delay, while the heuristic achieves lower energy consumptions. Larger fleets and less challenging schedules result in lower energy consumption and delays for all controllers.}
    \label{fig:energy}  
\end{figure}

Fig.~\ref{fig:energy} reports the pareto plot of delays vs total energy consumption per controller and fleet.
The more demanding schedules result in higher energy consumption for all controllers on average, with increases of 3.17~\unit{MWh} from the least demanding schedule on 02-03-2024, to 3.45~\unit{MWh} on 10-03-2024 and 3.65~\unit{MWh} on 07-11-2024.
Similarly, higher fleet sizes generally result in lower overall energy consumption, since vehicles reposition less to cover the same task set and consumption follows the distance driven.
The heuristic achieves the lowest energy consumptions, with the exception of the smallest fleet size on the most demanding schedule.
Similarly, the MCTS controller exhibits the highest consumption for the smallest fleet, but shows the best results with larger fleets. %
The MPC controller achieves the most consistent behavior in terms of energy consumption, performing better with small fleet sizes, but not improving as much as MCTS with larger fleets.
\subsubsection{Charging Patterns}
\begin{figure}[t]
    \centering
    \includegraphics[width=1\linewidth]{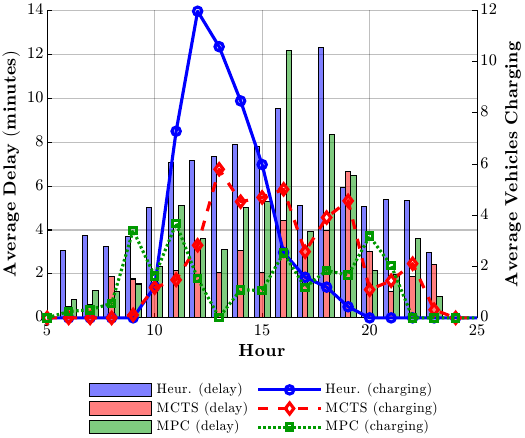}
    \caption{Average number of vehicles charging per hour for the Heuristic (blue), MPC (green), and MCTS (red) controllers overlayed on the average delay per hour across the day for the schedule of 02-03-2024 and a fleet of 25 vehicles. The heuristic charging pattern is concentrated always around midday, while MPC and MCTS spread charging preemptively and more evenly over time.}
    \label{fig:delaycharg2}  
\end{figure}

Fig.~\ref{fig:delaycharg2} exemplifies the charging strategies of each controller for the schedule of 02-03-2024 and a fleet of 25 vehicles. In the same plot, the average delay per hour across the day is overlayed to show the relation between charging patterns and delays.
The heuristic shows a concentrated charging pattern, consistent with the top-up when below $E^{\mathrm{min}}$ rule. It produces the highest peaks of simultaneous charging, with up to 12 average vehicles charging simultaneously around midday. This results in higher delays during these hours, as many vehicles are unavailable for task assignments while charging.
The MPC shows the most preemptive behavior, initiating charging sessions earlier in the day to avoid simultaneous charging sessions. It spreads charging over shorter charges, but still experiences some significant periods with delays.
The MCTS controller initiates charging at the same time as the heuristic, but spreads the sessions more evenly over time. It charges significantly more vehicles simultaneously than MPC, but avoids the delay peaks seen in the heuristic and MPC after 15:00.

Table~\ref{tab:combined-all} complements this, showing the average number of charging moments for each fleet and scenario. The heuristic charges between one and four times per vehicle, adding 54--84\% of $E^{\max}$ each time, whereas MPC and MCTS use shorter, partial top-ups of roughly 10--15\% of $E^{\max}$ (9.7--12.5\% and 12.3--14.9\%, respectively) and therefore charge more frequently.
In the case of smaller fleets, the number of charging events is significantly higher due to the increased need for repositioning and charging to cover all tasks with a lower number of available vehicles. In the case of the MPC and MCTS this leads to very complex charging operations to keep delays low, and signals the fleet design is approaching the minimum size required to cover the demand in a practical way.
Because it postpones to larger, deeper charges triggered by rule-based thresholds, the heuristic occupies 12 to 16 chargers simultaneously, compared with 6 to 10 for MPC and 8 to 11 for MCTS. Averaged over the nine scenarios this is 14.0 chargers for the heuristic against 8.0 for MPC and 9.3 for MCTS. A higher number of vehicles generally keeps constant or reduces the maximum number of chargers used, as charging sessions can be better distributed and there is more energy available in the fleet.
This is also reflected in the total energy charged by each controller. The heuristic charges the most energy overall, as it performs deeper rule-based charges that are based on fixed thresholds. The MPC charges the least energy, which is consistent with the penalization in the objective function. MCTS strikes a balance between both approaches, as the charging actions are based on thresholds but their duration and timing are optimized within the search process. These two latter approaches would rely on night charging to fully restore the fleet energy levels for the next day.
In terms of distance driven during the operations, the results vary significantly between controllers and fleet sizes. 
In general, higher fleet sizes result in lower distance travelled, as vehicles can be better allocated to nearby tasks without needing to reposition as often. However, the impact of the fleet size is stronger in the heuristic controller, and less significant in the MPC controller. 
The heuristic achieves the lowest average distance travelled with fleets 25 and 30, as it minimizes repositions for charging and it is exclusively based on distance to tasks and availability. With the smallest fleet size, however, it performs worse than MPC. 
The MCTS controller drives significantly more, as it repositions vehicles preemptively, resulting in increases of 36\% and 8.5\% over the heuristic and MPC, respectively, when pooled across all nine scenarios.
\begin{table*}[t]
  \centering
  \small
  \caption{Performance metrics for the controllers tested for each fleet size and schedule.}
  \label{tab:combined-all}
  \setlength{\tabcolsep}{4pt}
  \resizebox{\linewidth}{!}{%
  \begin{tabular}{ll|ccc|ccc|ccc|ccc|ccc|ccc}
    \toprule
    \multirow{2}{*}{Fleet} & \multirow{2}{*}{Controller}
      & \multicolumn{3}{c|}{Avg. Charging Events}
      & \multicolumn{3}{c|}{Avg. Charge [\%$E^{\max}$]}
      & \multicolumn{3}{c|}{Max chargers used}
      & \multicolumn{3}{c|}{Energy charged (MWh)}
      & \multicolumn{3}{c}{Total Distance (km)} \\
    \cmidrule(lr){3-5} \cmidrule(lr){6-8} \cmidrule(lr){9-11} \cmidrule(lr){12-14} \cmidrule(lr){15-17}
    &  & 02-03 & 10-03 & 07-11 & 02-03 & 10-03 & 07-11 & 02-03 & 10-03 & 07-11 & 02-03 & 10-03 & 07-11 & 02-03 & 10-03 & 07-11 \\
    \midrule
    \multirow{3}{*}{20}
      & Heuristic 
        & 2.80 & 2.70 & 3.75 
        & 62.1   & 68.3   & 54.2   
        & 15   & 14   & 15   
        & 3.13 & 3.32 & 3.66 
        & 372.56 & 454.57 & 582.68 \\
    
      & MPC 
        & 6.50 & 7.70 & 8.40 
        & 12.5   & 12.1   & 12   
        & 8   & 10   & 9   
        & 1.51 & 1.72 & 1.96 
        & 447.00 & 468.72 & 482.51 \\
    
      & MCTS 
        & 11.05 & 10.90 & 11.35 
        & 12.3   & 12.9   & 14.9   
        & 11   & 9   & 10   
        & 2.55 & 2.55 & 3.24 
        & 490.62 & 507.52 & 592.16 \\
    \addlinespace
    \multirow{3}{*}{25}
      & Heuristic 
        & 1.16 & 1.20 & 1.52 
        & 77.9   & 75.9   & 61.1   
        & 14   & 14   & 16   
        & 2.06 & 2.05 & 2.12 
        & 249.07 & 275.59 & 314.93 \\
    
      & MPC 
        & 4.12 & 4.88 & 6.00 
        & 11.4   & 11.9   & 11.3   
        & 6   & 9   & 7   
        & 1.12 & 1.41 & 1.69 
        & 332.83 & 383.94 & 462.70 \\
    
      & MCTS 
        & 5.24 & 6.80 & 8.48 
        & 14.4   & 13.4   & 13.8   
        & 9   & 9   & 10   
        & 1.73 & 2.16 & 2.75 
        & 370.73 & 422.87 & 558.70 \\
    \addlinespace
    \multirow{3}{*}{30}
      & Heuristic 
        & 1.10 & 1.03 & 1.07 
        & 77.1   & 81.0   & 83.9   
        & 13   & 13   & 12   
        & 2.31 & 2.27 & 2.44 
        & 210.55 & 234.09 & 276.28 \\
    
      & MPC 
        & 4.60 & 4.27 & 3.93 
        & 10.2   & 10.7   & 9.7   
        & 9   & 8   & 6   
        & 1.27 & 1.34 & 1.03 
        & 377.42 & 372.77 & 398.23 \\
    
      & MCTS 
        & 4.73 & 4.87 & 4.53 
        & 13.1   & 13.6   & 14.3   
        & 8   & 9   & 9   
        & 1.41 & 1.86 & 1.85 
        & 312.45 & 383.62 & 404.63 \\
    \bottomrule
  \end{tabular}%
  }
  \begin{minipage}{\linewidth}
    \footnotesize Note: The average number of charging events are per vehicle. The average energy charged is the average percentage of the battery charged per event. The maximum number of chargers used is the maximum number of chargers used at any one time during the day. Energy charged is the total energy charged by all vehicles during the schedule day. Total distance is the sum of all distance driven by all vehicles during the operation.
  \end{minipage}
\end{table*}

\section{Conclusion}\label{sec:Conclusion}
This paper compared three control strategies for a fleet of electric ground-support vehicles at Schiphol Airport, evaluated in a digital twin: (i) a nearest-vehicle heuristic, (ii) an MPC-based controller, and (iii) an MCTS-based controller.
We find MCTS achieves the lowest overall delays with low energy consumption, making it a promising approach for real-world deployment in EGSV fleets.
The heuristic controller yields reasonable solutions with minimal computational requirements, making it suitable for real-time applications with limited resources. Operations are also simplified, but at the expense of higher delays and infrastructure and fleet oversizing.
MPC offers the most balanced delay-energy trade-off for mid-sized fleets, but does not make substantial gains for the largest fleet sizes.
Larger fleets generally reduce charging infrastructure needs and energy consumption through more efficient vehicle assignments, a key upfront cost driver, but introduce greater operational complexity. 
Intelligent controllers are essential to capturing these efficiency gains and avoiding oversizing; smaller fleets benefit most, as they have less slack and are more vulnerable to inefficient heuristic use. 
Ensuring adequate KPI levels during this transition is critical, and is enabled by the use of DTs.
A key shortcoming of the present work is the lack of validation of the DT with recorded operational data; future work will address this and test the controllers across a wider variety of scenarios. 
Similarly, the use of manual task durations is a simplification that should be replaced by live occupancy data and upcoming mission information to produce more realistic estimates, to be developed in collaboration with airport stakeholders.
Extending this work to other GHOs, analyzing the performance of the controllers under constrained charging infrastructure, and further tuning of controller parameters are also promising avenues for future work, as is the development of adaptive control strategies that can learn from the DT environment and adjust their parameters in real time to optimize performance under changing conditions.

\newpage
\onecolumn

\appendices
\renewcommand{\thesection}{\arabic{section}}

\section{}\label{annex1}
\begin{table}[H]
\centering
\small
\begin{threeparttable}
\caption{OEM-based turnaround parameter table for incoming aircrafts in the DT.}
\label{tab:turnaround_oem_final}
\setlength{\tabcolsep}{4pt}
\begin{tabular*}{\textwidth}{@{\extracolsep{\fill}}ll>{\raggedleft\arraybackslash}p{1.15cm}*{6}{>{\raggedleft\arraybackslash}p{0.95cm}}@{}}
\toprule
Manufacturer & Model & MTOW (kg) & Deboard & Service & Board & Fuel & Unload & Load \\
\midrule
Airbus & A220-100 & 60,800 & 6.2 & 15.0 & 7.3 & 8.5 & 7.5 & 11.3 \\
Airbus & A220-300 & 67,600 & 6.4 & 15.0 & 9.1 & 9.4 & 8.3 & 13.3 \\

Airbus & A318-100 & 66,000 & 7.2 & 18.6 & 10.9 & 21.0 & 7.3 & 8.7 \\

Airbus & A319-100 & 76,500 & 8.2 & 18.6 & 12.3 & 21.0 & 5.3 & 6.2 \\
Airbus & A319neo & 75,500 & 8.2 & 18.6 & 12.3 & 21.0 & 5.3 & 6.2 \\

Airbus & A320-200 & 78,000 & 9.5 & 22.2 & 14.5 & 20.8 & 8.0 & 8.0 \\
Airbus & A320neo & 79,000 & 9.5 & 22.2 & 14.5 & 20.8 & 8.0 & 8.0 \\

Airbus & A321-200 & 93,500 & 9.0 & 25.8 & 17.4 & 16.0 & 9.5 & 9.5 \\
Airbus & A321neo & 93,500 & 9.0 & 25.8 & 17.4 & 16.0 & 9.5 & 9.5 \\
Airbus & A321neo-XLR & 101,000 & 10.0 & 22.0 & 19.2 & 32.4 & 9.5 & 9.5 \\

Airbus & A330-200 & 238,000 & 8.0 & 38.0 & 15.0 & 37.7 & 19.3 & 22.1 \\
Airbus & A330-300 & 238,000 & 9.0 & 41.0 & 17.0 & 40.0 & 24.1 & 27.7 \\
Airbus & A330-800 & 251,000 & 8.0 & 38.0 & 15.0 & 37.7 & 19.3 & 22.1 \\
Airbus & A330-900 & 251,000 & 9.0 & 41.0 & 17.0 & 40.0 & 24.1 & 27.7 \\

Airbus & A340-300 & 275,000 & 9.2 & 45.5 & 16.0 & 44.3 & 21.7 & 24.9 \\

Airbus & A350-900 & 280,000 & 9.8 & 42.5 & 17.5 & 51.3 & 21.7 & 24.9 \\
Airbus & A350-1000 & 322,000 & 10.2 & 39.5 & 22.0 & 51.3 & 24.1 & 27.7 \\

Airbus & A380-800 & 569,000 & 17.2 & 50.0 & 26.0 & 53.2 & 26.5 & 30.5 \\
\addlinespace

Embraer & EMB-120 ER & 11,990 & 3.0 & 6.0 & 3.0 & 4.0 & 3.0 & 3.0 \\
Embraer & EMB-135 LR & 20,000 & 4.0 & 5.0 & 4.5 & 12.0 & 4.5 & 6.0 \\
Embraer & ERJ-140 & 21,100 & 4.5 & 6.1 & 5.0 & 11.0 & 5.0 & 6.5 \\
Embraer & EMB-145 LR & 22,000 & 5.0 & 7.0 & 6.0 & 12.0 & 6.0 & 8.0 \\
Embraer & E170 LR & 37,200 & 3.9 & 6.0 & 5.4 & 8.5 & 3.4 & 5.5 \\
Embraer & E175 LR & 38,790 & 4.3 & 6.0 & 5.9 & 8.5 & 3.6 & 5.8 \\
Embraer & E190 LR & 50,300 & 5.1 & 7.0 & 7.1 & 11.7 & 4.3 & 7.1 \\
Embraer & E195 LR & 50,790 & 5.5 & 7.0 & 7.8 & 11.7 & 4.4 & 7.2 \\ %
\addlinespace

Bombardier & CRJ200 LR & 24,041 & 5.0 & 12.5 & 5.0 & 12.5 & 8.5 & 9.0 \\
Bombardier & CRJ700 & 32,999    & 6.0 & 12.5 & 6.0 & 6.2 & 9.5 & 10.0 \\
Bombardier & CRJ900 & 38,329    & 6.0 & 12.5 & 6.0 & 6.2 & 9.5 & 10.0 \\ %
\addlinespace %

Boeing & 737-100/500 & 60,554 & 5.0 & 20.0 & 5.0 & 7.0 & 15.0 & 15.0 \\
Boeing & 737-600 & 65,090 & 6.2 & 16.0 & 9.3 & 16.2 & 9.0 & 12.0 \\
Boeing & 737-700 & 70,080 & 7.8 & 18.0 & 11.7 & 16.2 & 7.0 & 13.0 \\
Boeing & 737-800 & 79,010 & 9.0 & 16.0 & 13.5 & 16.2 & 13.0 & 16.0 \\
Boeing & 737-900 & 85,139 & 10.0 & 17.0 & 15.0 & 16.2 & 14.0 & 18.0 \\

Boeing & 737-7     & 80,285 & 8.3 & 14.0 & 12.5 & 15.0 & 12.1 & 18.8 \\
Boeing & 737-8     & 82,191 & 9.0 & 14.8 & 13.5 & 15.0 & 13.1 & 20.3 \\
Boeing & 737-8-200 & 82,191 & 11.1 & 13.3 & 16.7 & 15.0 & 16.1 & 25.0 \\
Boeing & 737-9     & 88,300 & 14.0 & 14.0 & 18.0 & 15.0 & 17.3 & 26.9 \\
Boeing & 737-10    & 89,765 & 14.0 & 14.0 & 18.0 & 15.0 & 18.5 & 28.8 \\

Boeing & 747-400 & 396,893 & 11.0 & 45.0 & 18.0 & 28.0 & 24.0 & 24.0 \\
Boeing & 747-400ER & 412,770 & 11.0 & 45.0 & 18.0 & 53.0 & 24.0 & 24.0 \\

Boeing & 747-8 & 447,696 & 8.8 & 38.0 & 11.7 & 44.0 & 22.0 & 27.0 \\

Boeing & 757-200 & 115,666 & 10.0 & 20.0 & 21.0 & 15.0 & 15.0 & 17.0 \\
Boeing & 757-300 & 123,830 & 14.0 & 27.0 & 27.0 & 18.0 & 17.0 & 21.0 \\

Boeing & 767-200 & 142,880 & 7.4 & 22.0 & 8.6 & 24.0 & 12.0 & 18.0 \\
Boeing & 767-300 & 156,490 & 8.5 & 26.5 & 10.4 & 24.0 & 16.0 & 20.0 \\
Boeing & 767-400ER & 204,120 & 9.6 & 24.0 & 12.1 & 37.0 & 20.0 & 24.0 \\

Boeing & 777-200 & 233,600 & 8.3 & 31.0 & 12.0 & 23.0 & 18.0 & 23.0 \\
Boeing & 777-300 & 267,619 & 8.3 & 35.0 & 12.0 & 47.0 & 24.0 & 32.0 \\ %

Boeing & 777-200LR & 347,451 & 7.0 & 24.5 & 10.5 & 42.0 & 18.0 & 24.0 \\
Boeing & 777-300ER & 351,534 & 9.4 & 29.4 & 14.0 & 42.0 & 24.0 & 34.0 \\ %
Boeing & 777-F     & 347,814 & 0.0 & 10.0 & 0.0   & 58.0 & 12.0 & 18.0 \\ %

Boeing & 777-9 & 351,534 & 10.4 & 34.5 & 30.5 & 62.0 & 26.0 & 26.0 \\

Boeing & 787-8 & 227,930 & 8.9 & 19.5 & 11.7 & 35.0 & 16.0 & 22.0 \\
Boeing & 787-9 & 259,228 & 11.0 & 17.0 & 20.0 & 35.0 & 20.0 & 28.0 \\
Boeing & 787-10 & 260,362 & 12.3 & 20.0 & 21.7 & 35.0 & 22.0 & 32.0 \\

\bottomrule
\end{tabular*}
\begin{tablenotes}[flushleft]
\footnotesize
\item Notes: All values expressed in minutes unless otherwise noted. MTOW (Maximum weight for takeoff) is the maximum weight at the start of the takeoff run. Turnaround durations are extracted from manufacturer airport planning manuals and OEM turnaround-station charts. Actual turnaround times may vary based on specific airport operations, mission range and passenger load for the next flight, but have been assumed constant for this work.
\end{tablenotes}
\end{threeparttable}
\end{table}

\clearpage
\section{}\label{annex2}
\begin{table}[H]
\centering
\small
\begin{threeparttable}
\caption{Specifications and modelling assumptions of electric ground support vehicles used in the digital twin.}
\label{tab:vehicles}
\setlength{\tabcolsep}{3pt}
\begin{tabularx}{\textwidth}{@{}>{\raggedright\arraybackslash\hsize=0.8\hsize}X>{\raggedright\arraybackslash\hsize=1.35\hsize}X*{5}{>{\centering\arraybackslash\hsize=0.97\hsize}X}@{}}
\toprule
Variable & Description & Hydrant dispenser \cite{titan_hd150} & Baggage tractor \cite{mulag_comet4ehv} & Pushback tug \cite{goldhofer_ast2e_2025} & Service truck \cite{doll_xcatm} & Passenger bus \cite{man_lions_city_e_brochure} \\
\midrule
$m_\mathrm{v}$ (kg) & Vehicle mass & 7,500 & 4,500 & 16,965 & 11,250 & 15,950 \\
$E_\mathrm{max}$ (kWh) & Battery capacity & 90 & 50 & 160 & 100 & 400 \\
$P_\mathrm{C}$ (kW) & Charging power & 40 & 40 & 40 & 40 & 40\\
$m_\mathrm{c}$ (kg) & Cargo / passenger mass & 0 & 0 & variable\tnote{a} & 4,500 & 4,050\tnote{b} \\
$P^{\mathrm{tsk}}$ (kW) & Task power consumption & $\frac{q \cdot p}{3.6 \cdot 10^3 \cdot \eta}$ & 0 & As per~\eqref{eq:powerFx} with $m = m_\mathrm{c} + m_\mathrm{v}$ & 0 & 0 \\
\addlinespace
\multicolumn{7}{c}{\textit{Operational assumptions for pushback tugs}} \\
\midrule
$a_\mathrm{x,max}^\mathrm{p}$ (m/s$^2$) & Pushback acceleration & - & - & 1.0 & - & - \\
$v_\mathrm{p}$ (km/h) & Pushback speed & - & - & 14\tnote{c} & - & - \\

\addlinespace
\multicolumn{7}{c}{\textit{Operational assumptions for hydrant dispensers}} \\
\midrule
$q$ (m$^3$/h) & Pump flow rate & 150 & - & - & - & - \\
$p$ (Pa) & Pump pressure & $3.10 \cdot 10^5$ & - & - & - & - \\

\addlinespace
\multicolumn{7}{c}{\textit{Operational assumptions shared by all vehicles}} \\
\midrule
$\nu_\mathrm{tgt}$ (km/h) & Target driving speed & \multicolumn{5}{c}{25} \\
$a_\mathrm{x,max}$ (m/s$^2$) & Max. acceleration & \multicolumn{5}{c}{1.0} \\
$\rho$ (kg/m$^3$) & Air density & \multicolumn{5}{c}{1.225} \\
$\eta$ (-) & Powertrain efficiency & \multicolumn{5}{c}{0.8} \\
$C_\mathrm{d}$ (-) & Drag coefficient & \multicolumn{5}{c}{0.5} \\
$A$ (m$^2$) & Frontal area & \multicolumn{5}{c}{2.5} \\
$\mu$ (-) & Rolling resistance & \multicolumn{5}{c}{0.01} \\ %
\bottomrule
\end{tabularx}
\begin{tablenotes}[flushleft]
\footnotesize
\item
Notes: Values were agreed in consultation with Schiphol airport stakeholders and/or taken from OEM specifications where available.
[a] Varies according to the aircraft MTOW being serviced.
[b] Derived from a modelling assumption of 90 kg per passenger for 45 passengers.
[c] We assume that pushback always takes 3 minutes following \cite{Bubalo2017}.
\end{tablenotes}
\end{threeparttable}
\end{table}

\clearpage
\twocolumn

\section*{Acknowledgment}\label{sec:ack}
We thank Dr. O.~J.~Borsboom, F.~F.~Vehlhaber and Dr.~I.~New for proofreading this paper. This publication is part of the project GTD-Elektrifikatie, made possible in part by the Ministry of Economic Affairs and Climate Policy of the Netherlands. For the initial implementations of the ABM model and API pipeline we thank Vincent Kuipers and Thomas Leniere.

\printbibliography

\end{document}